\documentclass[epjST]{svjour}

\usepackage[utf8]{inputenc}
\usepackage{graphics,graphicx}
\usepackage{amssymb,amsfonts,amsmath,bm}
\usepackage[colorlinks=true,citecolor=blue,urlcolor=blue]{hyperref}
\usepackage{cite}
\usepackage{units}

\def \beq {\begin{equation}}
\def \edq {\end{equation}}

\def \calt {{\cal{T}}}

\def \calr {{\cal{R}}}

\def\Circlearrowleft{\ensuremath{%
  \rotatebox[origin=c]{180}{$\circlearrowleft$}}}
\def\Circlearrowright{\ensuremath{%
  \rotatebox[origin=c]{180}{$\circlearrowright$}}}

\begin{document}
\title{Phase-coherent caloritronics with ordinary and topological Josephson junctions}
\titlerunning{Phase-coherent caloritronics with ordinary and topological Josephson junctions}

\author{
Sun-Yong Hwang\thanks{\href{mailto:sunyong@thp.uni-due.de}{sunyong@thp.uni-due.de}} \and
Bj\"orn Sothmann\thanks{\href{mailto:bjoerns@thp.uni-due.de}{bjoerns@thp.uni-due.de}}
}
\institute{Theoretische Physik, Universit\"at Duisburg-Essen and CENIDE, D-47048 Duisburg, Germany}
\date{\today}

\abstract{
We provide a brief and comprehensive overview over recent developments in the field of phase-coherent caloritronics in ordinary and topological Josephson junctions.
We start from the simple case of a short, one-dimensional superconductor-normal metal-superconductor (S-N-S) Josephson junction and derive the phase-dependent thermal conductance within the Bogoliubov-de Gennes formalism. Then, we review the key experimental breakthroughs that have triggered the recent growing interest into phase-coherent heat transport. They include the realization of thermal interferometers, diffractors, modulators and routers based on superconducting tunnel junctions. Finally, we discuss very recent theoretical findings based on superconductor-topological insulator-superconductor (S-TI-S) Josephson junctions that show interesting heat transport properties due to the interplay between topological band structures and superconductivity.
} %end of abstract
\maketitle
\section{Introduction}
The discovery of macroscopic thermoelectric effects and a proper understanding of heat transport date back by almost two centuries~\cite{fourier_theorie_1822,seebeck_ueber_1826}. Recently, the interest in these phenomena in the quantum regime has been reignited starting with the pioneering work of Hicks and Dresselhaus~\cite{hicks_effect_1993,hicks_thermoelectric_1993} who realized that nanostructured materials can outperform classical thermoelectric materials.
One of the main reasons for this boost at the nanoscale is the envisaged great enhancement of energy harvesting efficiency with a careful engineering of nanostructures. Mahan and Sofo theoretically showed that the ideal efficiency can be achieved with sharp spectral features~\cite{mahan_best_1996} that arise, e.g., in quantum dots with discrete energy levels~\cite{staring_coulomb-blockade_1993}. It was experimentally found that decreasing the dimension of the system can dramatically enhance the thermoelectric efficiency~\cite{venkatasubramanian_thin-film_2001}. Furthermore, quantum thermoelectrics and heat transport are intimately connected to a number of fundamental physics problems in the context of quantum thermodynamics and thermometry~\cite{schwab_measurement_2000,meschke_single-mode_2006,jezouin_quantum_2013,dutta_thermal_2017,ronzani_tunable_2018,wang_fast_2018,maillet_optimal_2019}. For comprehensive recent reviews, we refer the readers to Refs.~\cite{shakouri_recent_2011,radousky_energy_2012,kosloff_quantum_2014,sanchez_focus_2014,sothmann_thermoelectric_2015,pekola_towards_2015,sanchez_nonlinear_2016,benenti_fundamental_2017,pekola_thermodynamics_2019}.

In order to maximize the thermoelectric efficiency, one usually wants to maximize the electrical conductance while at the same time minimizing the thermal conductance. The latter can be achieved, e.g., by reducing the number of transport channels that are not thermoelectrically active such as phonons. This reduction can be accomplished by introducing suitable interfaces that scatter phonons or thermoelectric generators based on nanowires with a limited number of available phonon modes \cite{rowe_CRC_1995,beekman_better_2015,bourgeois_reduction_2016}.
However, even when phononic channels are completely eliminated, the minimal electronic thermal conductance that can be achieved at a given electrical conductance is often limited by the Wiedemann-Franz law.

As an alternative route, one can actively exploit and manage the heat which is unavoidably present in electronic nanostructures. Very recently, a great deal of progress has been made in this direction with the introduction of so called phase-coherent caloritronics in superconducting circuits~\cite{fornieri_towards_2017}. The key idea here is to achieve coherent control over heat flows in superconducting hybrid systems where particular use is made of macroscopic quantum coherence inherent to the superconducting phase difference in a Josephson junction~\cite{josephson_possible_1962}.
In such a junction, heat is carried by quasiparticles with energies above the superconducting gap. Importantly, they can be transmitted across the junction in normal tunneling events as well as in Andreev-like processes where an electronlike quasiparticle is converted into a holelike quasiparticle and vice versa together with the creation or annihilation of a Cooper pair in the condensate. It is these latter processes that give rise to a phase-dependent contribution to the thermal conductance of a Josephson junction as first predicted for tunnel junctions~\cite{maki_entropy_1965,maki_entropy_1966,guttman_phase-dependent_1997,guttman_interference_1998,golubev_heat_2013}, weak links~\cite{zhao_phase_2003}, and point contacts~\cite{zhao_heat_2004}. More theoretical works on phase-dependent heat transport have been recently carried out in ferromagnetic Josephson junctions~\cite{giazotto_phase-tunable_2013,bergeret_phase-dependent_2013}, ac-driven systems~\cite{virtanen_thermal_2014}, heat current fluctuations~\cite{virtanen_fluctuation_2015}, dephasing of flux qubits~\cite{spilla_measurement_2014}, quantum dot hybrids~\cite{kamp_phase-dependent_2019}, diffusive junctions~\cite{hajiloo_mesoscopic_2019}, quantum point contacts~\cite{pershoguba_thermopower_2019}, solitons on long Josephson junctions~\cite{guarcello_solitonic_2018,guarcello_phase-coherent_2018,guarcello_solitonic_2018-1} and unconventional superconductors and superfluids~\cite{bauer_phase-dependent_2019}.

While phase-dependent heat currents have been predicted more than half a century ago~\cite{maki_entropy_1965,maki_entropy_1966}, they have been measured experimentally only very recently~\cite{giazotto_josephson_2012}. This experimental breakthrough has then triggered the conception and realization of various caloritronic devices including thermal interferometers~\cite{martinez-perez_fully_2013,fornieri_nanoscale_2016,fornieri_0-pi_2017}, heat transistors~\cite{giazotto_proposal_2014,fornieri_negative_2016}, diodes~\cite{martinez-perez_efficient_2013,giazotto_thermal_2013,fornieri_normal_2014,martinez-perez_rectification_2015}, switches~\cite{sothmann_high-efficiency_2017}, thermal memory~\cite{guarcello_josephson_2018}, and refrigerators~\cite{solinas_microwave_2016,hofer_autonomous_2016,vischi_thermodynamic_2019}.

At the same time, Josephson junctions based on topological insulators (TIs) have attracted an enormous interest recently, with the latter being able to exhibit intriguing thermoelectric and heat transport properties\cite{hwang_nonlinear_2014,vannucci_interference-induced_2015,ronetti_spin-thermoelectric_2016,ronetti_polarized_2017,roura-bas_enhanced_2018,bohling_thermoelectric_2018,roura-bas_helical_2018,aharon-steinberg_phenomenological_2019}. This is due to the possibility to create topologically nontrivial superconducting states in such junctions which can host exotic surface states such as Majorana bound states~\cite{alicea_new_2012,beenakker_search_2013,aguado_majorana_2017}. In particular, surface states of a topological insulator can give rise to helical Andreev bound states with a zero-energy crossing at phase difference $\phi=\pi$~\cite{fu_superconducting_2008}. These degenerate zero-energy states are topologically protected, i.e. in contrast to Andreev bound states in a S-N-S junction they do not split in the presence of backscattering as long as time-reversal symmetry is preserved. The crossing of Andreev bound state in turn gives rise to a Josephson current that is $4\pi$-periodic rather than exhibiting the usual $2\pi$-periodicity~\cite{fu_josephson_2009}. This fractional Josephson effect has been observed in a series of recent experiments~\cite{rokhinson_fractional_2012,wiedenmann_4pi-periodic_2016,bocquillon_gapless_2017,deacon_josephson_2017,li_4pi-periodic_2018,laroche_observation_2019} despite challenges arising from quasiparticle poisoning and additional, topologically trivial modes.
Very recently, phase-dependent heat transport has been suggested as an alternative way of probing the existence of topological Andreev bound states in Josephson junctions~\cite{sothmann_fingerprint_2016,sothmann_high-efficiency_2017,bours_phase-tunable_2019,bauer_phase-dependent_2019}.

This review is organized as follows. In section \ref{SNS}, we consider a short, one-dimensional S-N-S junction and derive the phase-dependent thermal conductance within the Bogoliubov-de Gennes formalism. Section \ref{Exp} covers the key experiments based on direct current superconducting quantum interference devices (dc SQUIDs) which pave the way for phase-coherent caloritronics. These experiments are realizations of heat interferometers, diffractors, modulators, phase-controllers and routers. In section \ref{STIS}, we give a summary of the several recent theoretical works based on the S-TI-S junctions. A recent theoretical proposal for a phase-coherent heat circulator based on the multiterminal S-N-S setup is also discussed. Final conclusions are provided in section \ref{sum}.

\section{S-N-S junctions}\label{SNS}
 We consider thermal transport by quasiparticles above the superconducting gap in a one-dimensional S-N-S Josephson junction where the superconducting coherence length is much larger than the junction length $L$ such that we can focus on the limit $L\to0$.
In order to describe heat transport in such a junction, we make use of the Bogoliubov-de Gennes approach where the number of degrees of freedom is doubled to account properly for the particle-hole symmetry of a superconductor~\cite{blonder_transition_1982}. With a basis chosen by spin-up electrons and spin-down holes, we can write down the Bogoliubov-de Gennes equation
\begin{equation}
	H_{\text{BdG}}\Psi_{\gamma}^{e/h}(x)=\omega\Psi_{\gamma}^{e/h}(x),
\end{equation}
with the Bogoliubov-de Gennes Hamiltonian given by
\beq\label{eq:HBdG1}
H_{\text{BdG}}=\begin{pmatrix}
h(x) & \Delta(x)\\
\Delta^*(x) & -h^*(x)
\end{pmatrix},
\edq
where
\beq\label{eq:h1}
h(x)=-\frac{\hbar^2\partial_x^2}{2m}-\mu
\edq
denotes the single-particle Hamiltonian that describes noninteracting electrons with mass $m$ and electrochemical potential $\mu$. The hole Hamiltonian is given by $-h^*(x)$.
Electrons and holes are coupled by the superconducting order parameter which in a minimal model can be written as
\beq
\Delta(x)=\Delta e^{i\phi_\text{L}}\Theta(-x)+\Delta e^{i\phi_\text{R}}\Theta(x).
\edq
Here, we have assumed that the superconducting order parameter has the same absolute value $\Delta$ on both sides of the junction. Furthermore, we assume an abrupt change of the order parameter at the interface, i.e., we neglect any potential (inverse) proximity effect which is a reasonable assumption for a one-dimensional junction~\cite{beenakker_josephson_1991,beenakker_universal_1991}. With the above choice of order parameters, the phase bias between two superconducting leads is given by $\phi=\phi_\text{L}-\phi_\text{R}$.

The eigenfunctions of the Bogoliubov-de Gennes equation are given by electronlike ($e$) and holelike ($h$) states in the two superconductors $\gamma=\text{L,R}$, respectively:
\begin{subequations}\label{eq:waveSC}
\begin{align}
\Psi_{\gamma,\pm}^e(x)=\begin{pmatrix}u\\ve^{-i\phi_\gamma}\end{pmatrix}e^{\pm i k_F x},\\
\Psi_{\gamma,\pm}^h(x)=\begin{pmatrix}v\\ue^{-i\phi_\gamma}\end{pmatrix}e^{\pm i k_F x},
\end{align}
\end{subequations}
where $\pm$ denotes the propagating direction along the $x$ axis and we have made use of the Andreev approximation\cite{beenakker_universal_1991} which is valid when the electrochemical potential $\mu\gg \Delta,\omega$ is the largest energy scale in the problem such that electronlike and holelike quasiparticles have identical momenta $\hbar k_F=\sqrt{2m\mu}$. The coherence factors entering the eigenfunctions are given by
\begin{subequations}
\begin{align}
u=\sqrt{\frac{1}{2}\left(1+\frac{\sqrt{\omega^2-\Delta^2}}{\omega}\right)},\\
v=\sqrt{\frac{1}{2}\left(1-\frac{\sqrt{\omega^2-\Delta^2}}{\omega}\right)}.
\end{align}
\end{subequations}
In order to model scattering at the interface between the two superconductors, we introduce a delta-barrier potential $V(x)=U\delta(x)$. Scattering at the junction can then conventiently be characterized by the barrier strength
\beq
Z=\frac{Um}{\hbar^2k_F}
\edq
which is related to the transmission probability of the junction in the normal state via
\beq
\tau=\frac{1}{1+Z^2}.
\edq
In order to determine the transmission probability through the junction, we solve the scattering problem by matching the wavefunctions at the boundary for the case of an incoming electronlike quasiparticle that gives rise to reflected and transmitted electronlike and holelike quasiparticles. The wavefunctions in the left and right superconductors can then be written as
\begin{subequations}\label{eq:psiLR}
\beq
\Psi_\text{L}(x)=\begin{pmatrix}u\\ve^{-i\phi_\text{L}}\end{pmatrix}e^{ik_Fx}
+r_{ee}\begin{pmatrix}u\\ve^{-i\phi_\text{L}}\end{pmatrix}e^{-ik_Fx}
+r_{he}\begin{pmatrix}v\\ue^{-i\phi_\text{L}}\end{pmatrix}e^{ik_Fx},
\edq
\beq
\Psi_\text{R}(x)=t_{ee}\begin{pmatrix}u\\ve^{-i\phi_\text{R}}\end{pmatrix}e^{ik_Fx}
+t_{he}\begin{pmatrix}v\\ue^{-i\phi_\text{R}}\end{pmatrix}e^{-ik_Fx},
\edq
\end{subequations}
where $r_{ee}$, $r_{he}$, $t_{ee}$, and $t_{he}$ denote the scattering amplitudes of the reflected and transmitted electronlike and holelike quasiparticles. They can be evaluated from the boundary conditions
\begin{subequations}
\begin{align}
&\Psi_\text{R}(0)=\Psi_\text{L}(0),\\
&\partial_x\Psi_\text{R}(0)-\partial_x\Psi_\text{L}(0)=2Zk_F\Psi_\text{R,L}(0),
\end{align}
\end{subequations}
to Eq. \eqref{eq:psiLR}. One then finds
\begin{align}
&\calt_{ee}(\omega,\phi)=|t_{ee}|^2
=\frac{\tau(\omega^2-\Delta^2)(\omega^2-\Delta^2\cos^2\frac{\phi}{2})}
	{\left[\omega^2-\Delta^2(1-\tau\sin^2\frac{\phi}{2})\right]^2},\\
&\calt_{he}(\omega,\phi)=|t_{he}|^2
=\frac{\tau(1-\tau)(\omega^2-\Delta^2)\Delta^2\sin^2\frac{\phi}{2}}
	{\left[\omega^2-\Delta^2(1-\tau\sin^2\frac{\phi}{2})\right]^2}.
\end{align} 
The corresponding transmission amplitudes for an incoming holelike quasiparticle are given by $\calt_{hh}(\omega,\phi)=\calt_{ee}(\omega,\phi)$ and $\calt_{eh}(\omega,\phi)=\calt_{he}(\omega,\phi)$. Thus, the total transmission function $\calt(\omega,\phi)=\calt_{ee}(\omega,\phi)+\calt_{eh}(\omega,\phi)+\calt_{hh}(\omega,\phi)+\calt_{he}(\omega,\phi)$ reads
\beq\label{eq:TSNS}
\calt(\omega,\phi)=\frac{2\tau(\omega^2-\Delta^2)\left[\omega^2-\Delta^2(\cos\phi+\tau\sin^2\frac{\phi}{2})\right]}
	{\left[\omega^2-\Delta^2(1-\tau\sin^2\frac{\phi}{2})\right]^2}.
\edq
It gives rise to the phase-dependent thermal conductance
\beq\label{eq:kappaSNS}
\kappa(\phi)=\frac{1}{h}\int d\omega\, \omega\calt(\omega,\phi)\frac{df}{dT},
\edq
where $f=[1+\exp(\omega/k_BT)]^{-1}$ is the Fermi-Dirac distribution function at equilibrium.
Therefore, the thermal conductance of the junction depends on the superconducting phase bias $\phi=\phi_\text{L}-\phi_\text{R}$ and can be controlled by it.
\begin{figure}
	\centering
	\includegraphics[width=.6\textwidth]{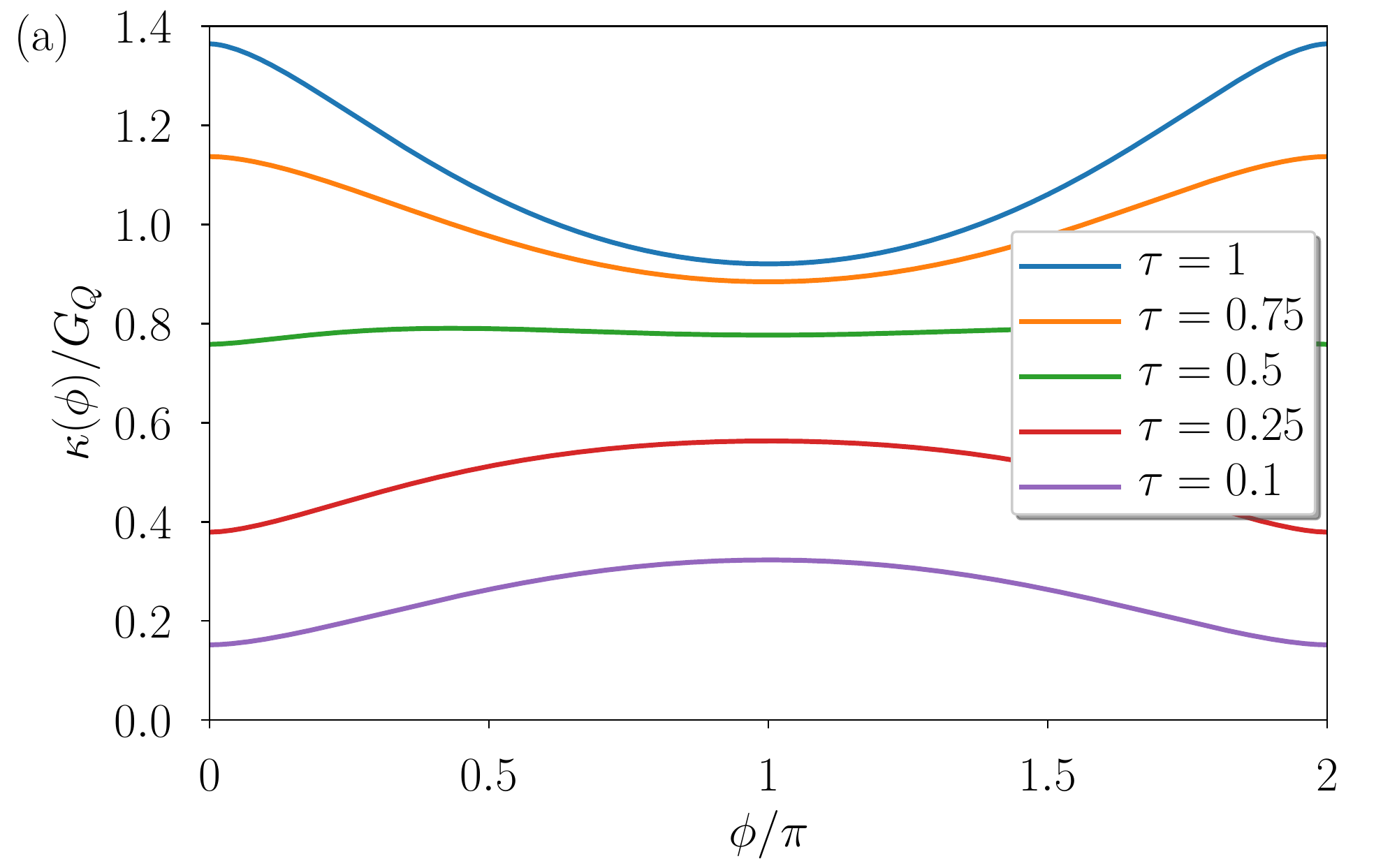}\\
	\includegraphics[width=.6\textwidth]{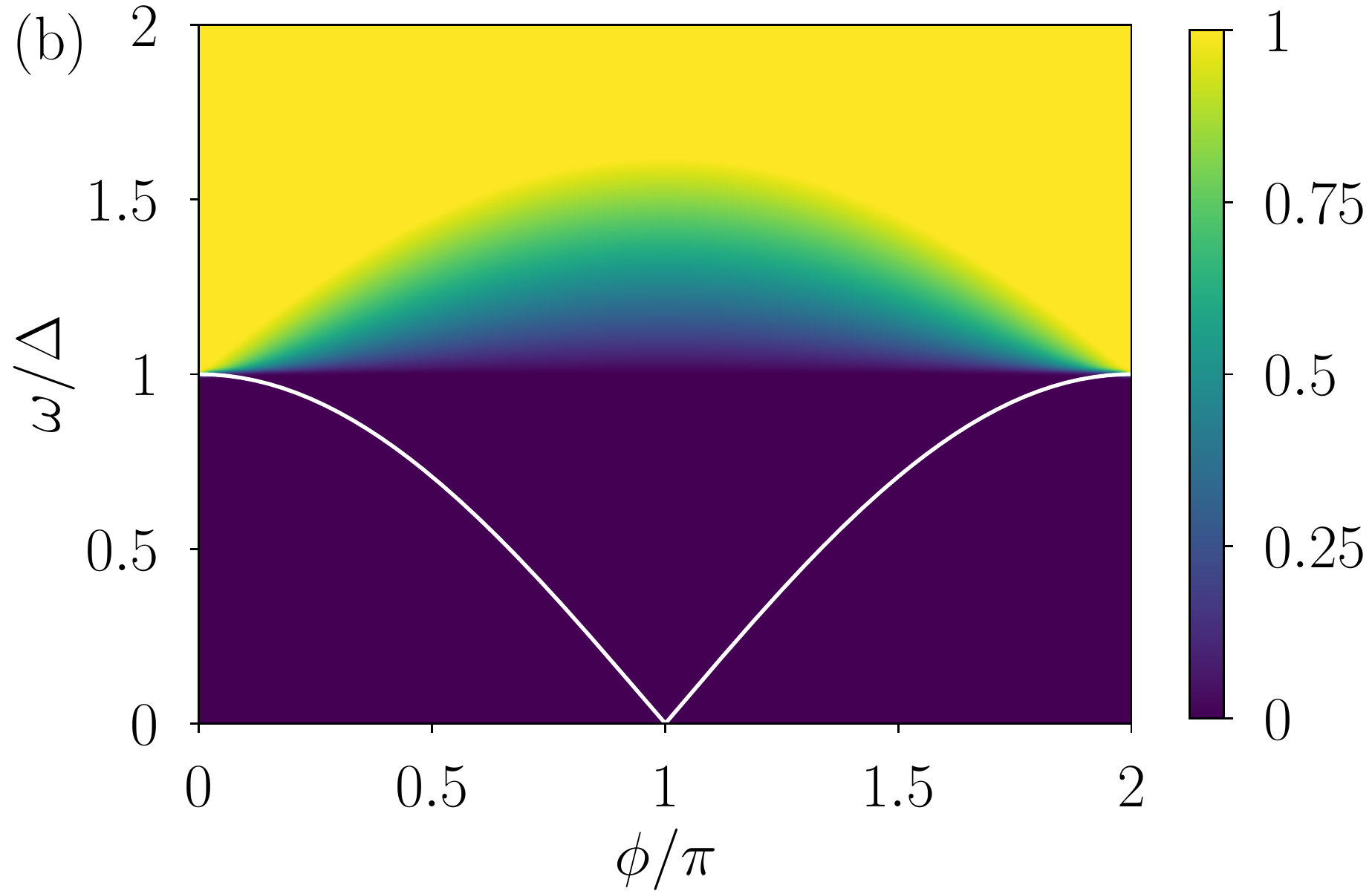}\\
	\includegraphics[width=.6\textwidth]{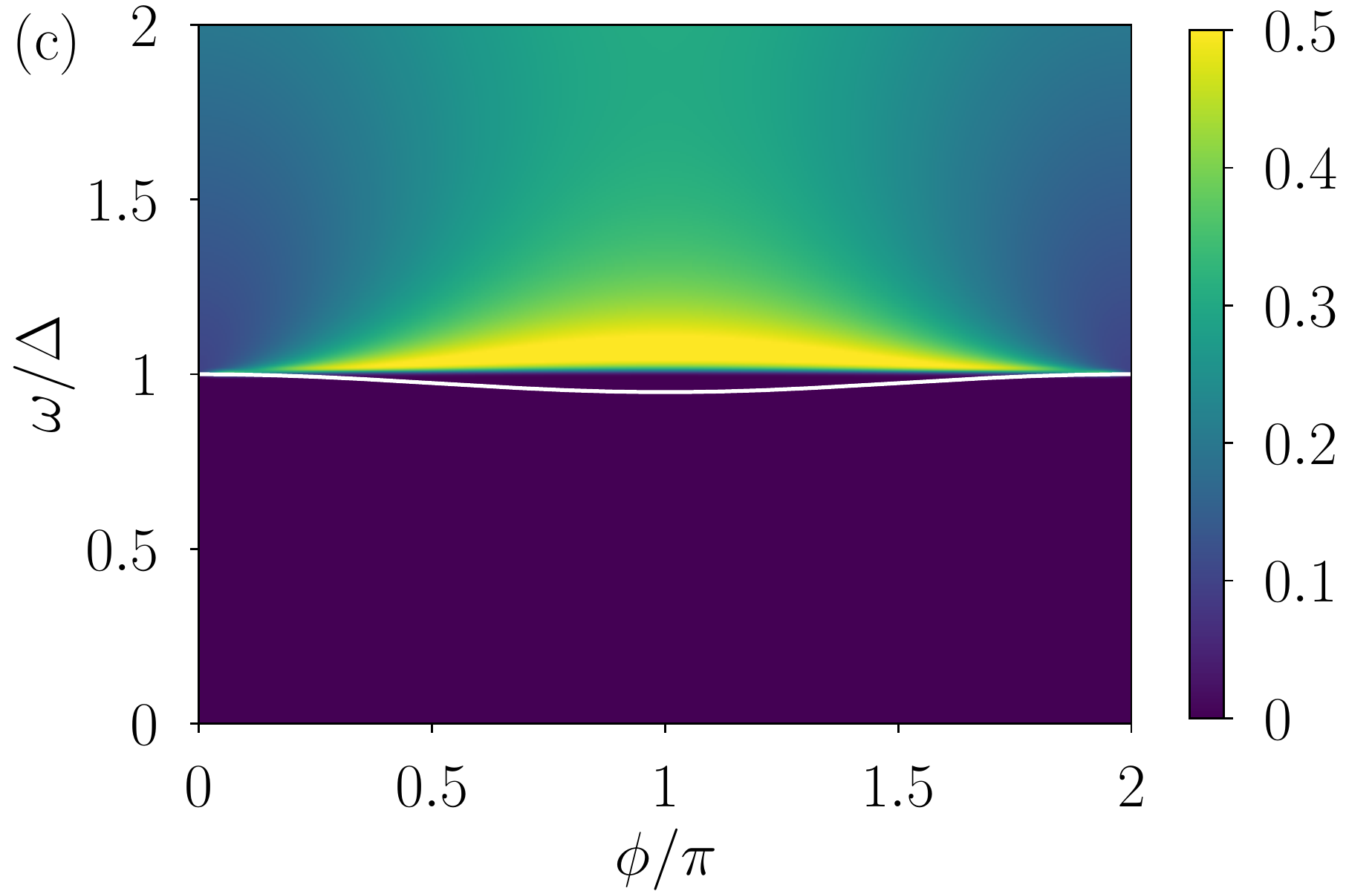}
	\caption{\label{fig:1dSNS}(a) Phase-dependent thermal conductance $\kappa(\phi)$ in units of the thermal conductance quantum $G_Q=\pi^2k_B^2T/(3h)$ for a short one-dimensional S-N-S junction with different normal state transmissions. Temperature is $k_BT=\Delta/2$. The transmission probability for quasiparticles is shown in (b) for $\tau=1$ and (c) for $\tau=0.1$. White lines inside the superconducting gap indicate the energy of Andreev bound states.}
\end{figure}
In Fig.~\ref{fig:1dSNS}(a), we show the phase-dependent thermal conductance for different transparencies of the junction in the normal state. For the case of a transparent junction, $\tau=1$, the thermal conductance exhibits a maximum at phase difference $\phi=0$ and a minimum at $\phi=\pi$. This can be traced back to the phase-dependent transmission shown in Fig.~\ref{fig:1dSNS}(b). As the Andreev bound state approaches zero energy for phase difference $\pi$ it gives rise to a strongly reduced transmission through the junction which is reflected directly in the thermal conductance. When the junction transparency is reduced, the thermal conductance decreases as expected. In addition, the phase dependence in the tunneling limit is qualitatively different and exhibits a maximum at $\phi=\pi$ rather than a minimum. This change in behavior can again be understood by analyzing the transmission function shown in Fig.~\ref{fig:1dSNS}(c). For $\phi=\pi$, the Andreev bound state stays close to the superconducting gap. It thus gives rise to a resonant transmission for energies slightly above the superconducting gap which leads to an increased thermal conductance.

The above discussion of phase-dependent thermal transport applies to ballistic Josephson junctions with a single conducting channel. Recently, heat transport has also been investigated in diffusive Josephson junctions that contain a large number of transport channels~\cite{hajiloo_mesoscopic_2019}. It was found that the phase-dependent contribution to the thermal conductance is eliminated on average. At the same time, it was shown that thermal conductance fluctuations as well as weak-localization corrections to the thermal conductannce still are still phase-sensitive quantities.

\section{Experimental progress}\label{Exp}
While the existence of phase-dependent heat currents has been predicted more than half a century ago~\cite{maki_entropy_1965,maki_entropy_1966}, their experimental detection has for a long time been hindered by technical difficulties in measuring heat currents. This has changed only very recently due to tremendous progress in nanoscale thermometry~\cite{giazotto_opportunities_2006} and has led to the first experimental evidence of phase-coherent heat transport in 2012~\cite{giazotto_josephson_2012}. This ground-breaking experiment has been followed by a series of other interesting experiments which we will shortly review in the following.

\subsection{Josephson heat interferometers}
\begin{figure}
\centering
\begin{minipage}{.3\textwidth}
\includegraphics[width=\textwidth]{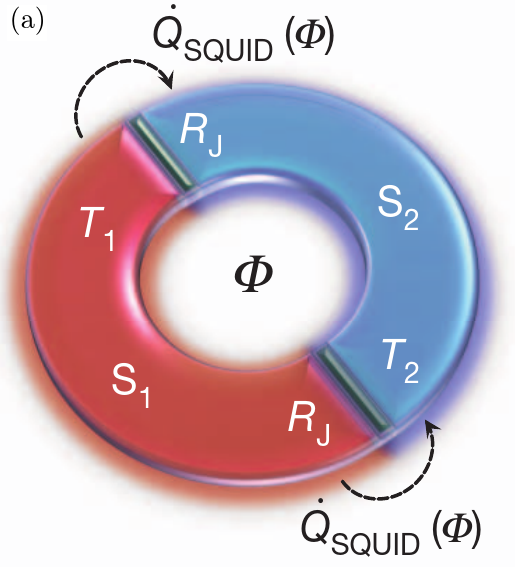}
\end{minipage}
\begin{minipage}{.4\textwidth}
\includegraphics[width=\textwidth]{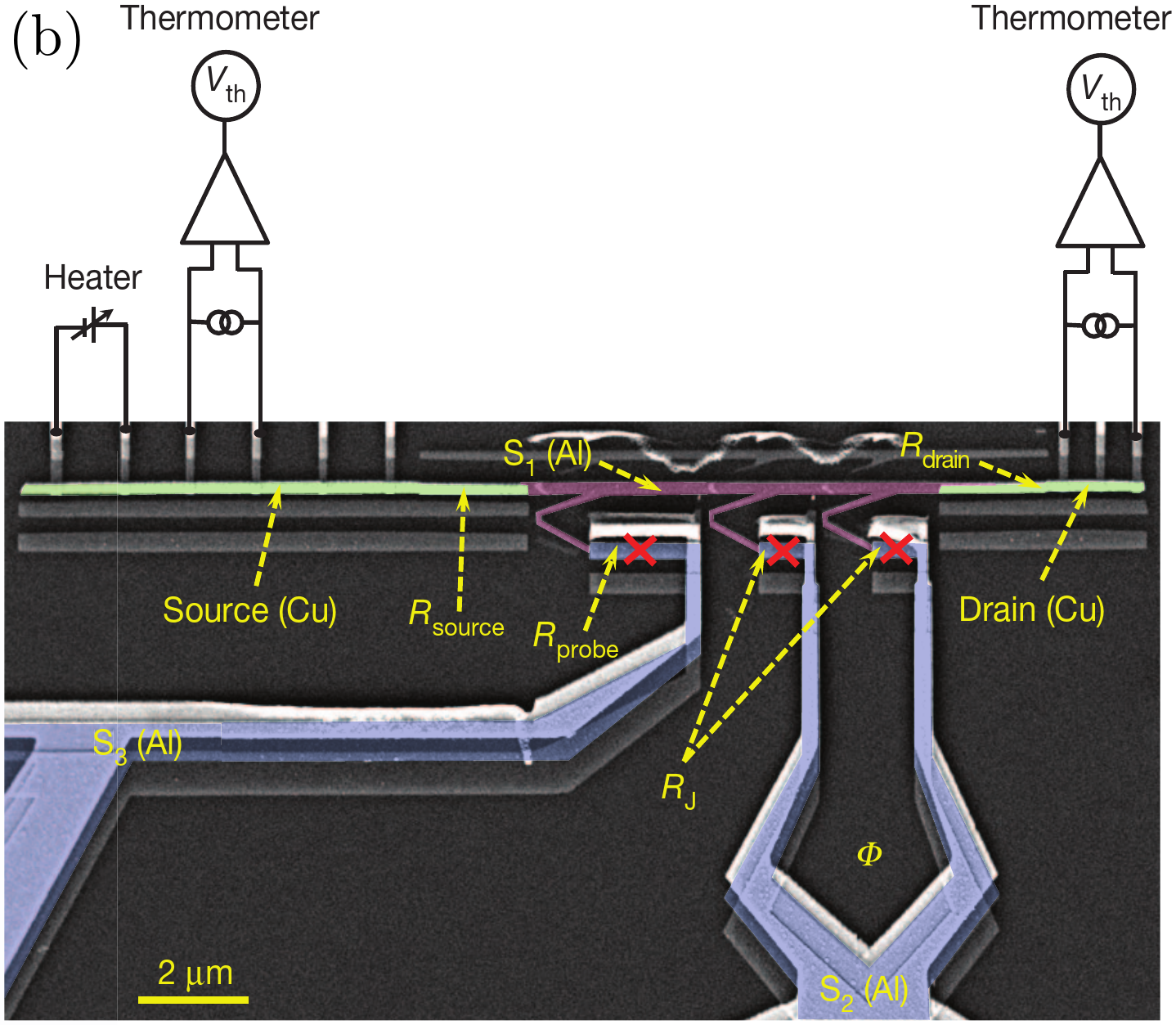}
\end{minipage}
\begin{minipage}{.27\textwidth}
\includegraphics[width=\textwidth]{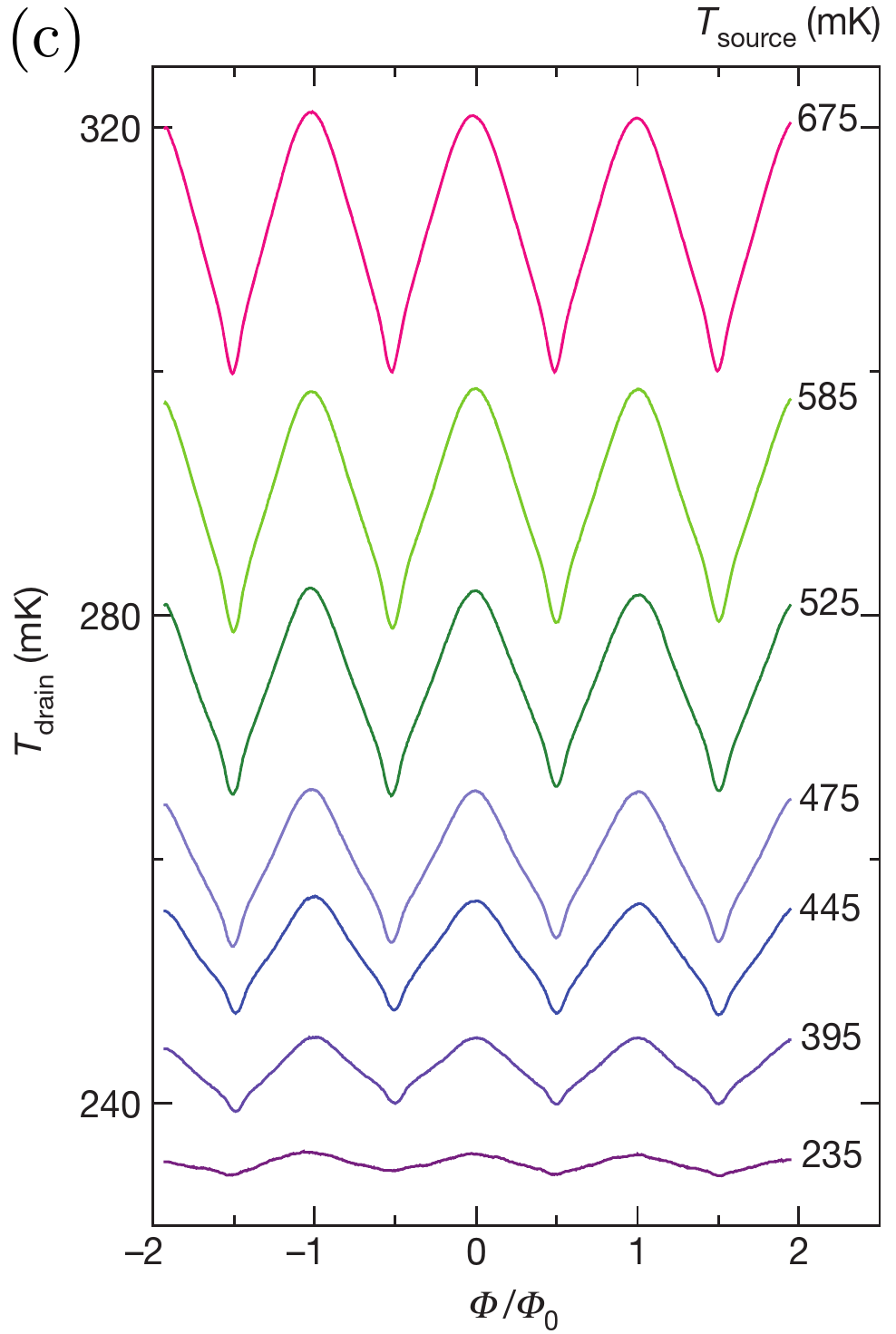}
\end{minipage}
\caption{(a) A direct current SQUID with two identical superconductors $S_1$ and $S_2$ residing at different temperatures $T_1$ and $T_2$. $R_J$ is the normal state resistance. (b) A false color image of the implementation of the Josephson heat interferometer. A source (Cu) electrode is coupled to a heater for raising the temperature of the source and also to a thermometer for measuring the temperature. Temperature variation of the drain (Cu) electrode is monitored by an attached thermometer which indicates the transported heat currents. Three Josephson junctions comprised of superconducting Al electrodes are marked with red crosses where the heat interference can be controlled by the magnetic flux $\Phi$. (c) Measured temperature variation of the drain as a function of the magnetic flux. Adapted from Ref.~\cite{giazotto_josephson_2012}.
}
\label{fig:nature}
\end{figure}

In Ref.~\cite{giazotto_josephson_2012}, the first evidence of phase-dependent thermal transport was reported. A Josephson heat interferometer realized with a symmetric dc SQUID, see Fig. \ref{fig:nature}(a), is temperature biased $T_1>T_2$ giving rise to the stationary heat flow through each of the junctions
\beq\label{eq:Qsquid}
\dot Q_\text{SQUID}(\Phi)=\dot Q_\text{QP}-\dot Q_\text{AR}\left|\cos\left(\frac{\pi\Phi}{\Phi_0}\right)\right|,
\edq
where $\Phi$ is the magnetic flux through the SQUID and $\Phi_0=h/(2e)$ denotes the magnetic flux quantum.
Importantly, $\dot Q_\text{AR}(T_1,T_2)$ arises from Andreev-like processes where an electronlike quasiparticle is converted into a holelike particle while creating or annihilating a Cooper pair and, thus, carries information on the superconducting phase while $\dot Q_\text{QP}(T_1,T_2)$ comes from the flux-independent quasiparticle heat flow. $\dot Q_\text{AR}(T_1,T_2)$ vanishes if at least one of the superconductors is in the normal state confirming that this term originates from superconducting phase coherence. Furthermore, when $T_1=T_2$ it follows that $\dot Q_\text{QP}=\dot Q_\text{AR}=0$ in Eq. \eqref{eq:Qsquid} indicating that the effect is due to the temperature bias. Finally, we remark that we always have $\dot Q_\text{QP}>\dot Q_\text{AR}$, i.e., heat flows from hot to cold in agreement with the second law of thermodynamics.
The experimental realization of the Josephson heat interferometer is shown in Fig.~\ref{fig:nature}(b). It consists of a source and a drain Cu electrodes which are connected by several superconducting contacts that serve as a heater and as a thermometer. In between the Cu electrodes, several superconducting Al electrodes define three Josephson junctions marked with red crosses in Fig.~\ref{fig:nature}(b). The phase difference across two of these junctions can be controlled by an external magnetic field. The key idea of the experiment is to heat the source electrode to $T_\text{source}$ and measure the temperature of the drain electrode $T_\text{drain}$ as the phase difference across the Josephson junctions is modulated. The temperature variation can be connected to the heat currents in the setup via a simple thermal model that accounts for all heat currents flowing in and out of the superconductor $S_1$ and the drain, respectively,
\begin{subequations}\label{eq:heatbalance}
\begin{align}
\dot Q_\text{source}=\dot Q_\text{probe}+\dot Q_\text{SQUID}(\Phi)+\dot Q_\text{drain},\\
\dot Q_\text{drain}=\dot Q_\text{e-p,drain}.
\end{align}
\end{subequations}
Here, $\dot Q_\text{source}$ denotes the heat flowing from the source into $S_1$. Similarly, $\dot Q_\text{drain}$ is the heat flowing between $S_1$ and the drain electrode. $\dot Q_\text{probe}$ is the heat current flowing from $S_1$ into the probe terminal and $\dot Q_\text{e-p,drain}$ describes heat losses into the substrate via electron-phonon couplings in the drain.

The experimentally measured temperature of the drain electrode shown in Fig.~\ref{fig:nature}(c) shows indeed the expected periodic modulation with the magnetic flux $\Phi$. Furthermore, the experimental results are in good agreement with the theoretical predictions by Maki and Griffin~\cite{maki_entropy_1965,maki_entropy_1966}. As shown, increasing $T_\text{source}$ monotonically enhances the average drain temperature $T_\text{drain}$. However, it has been observed that the amplitude of the modulation initially increases and then saturates with larger $T_\text{source}$.
% The experiment paved the way for the newly emerging field where one manipulates heat in a phase-coherent manner with various caloritronic devices~\cite{fornieri_towards_2017}. 

\begin{figure}
\centering
\begin{tabular}{c}
\includegraphics[width=0.6\textwidth]{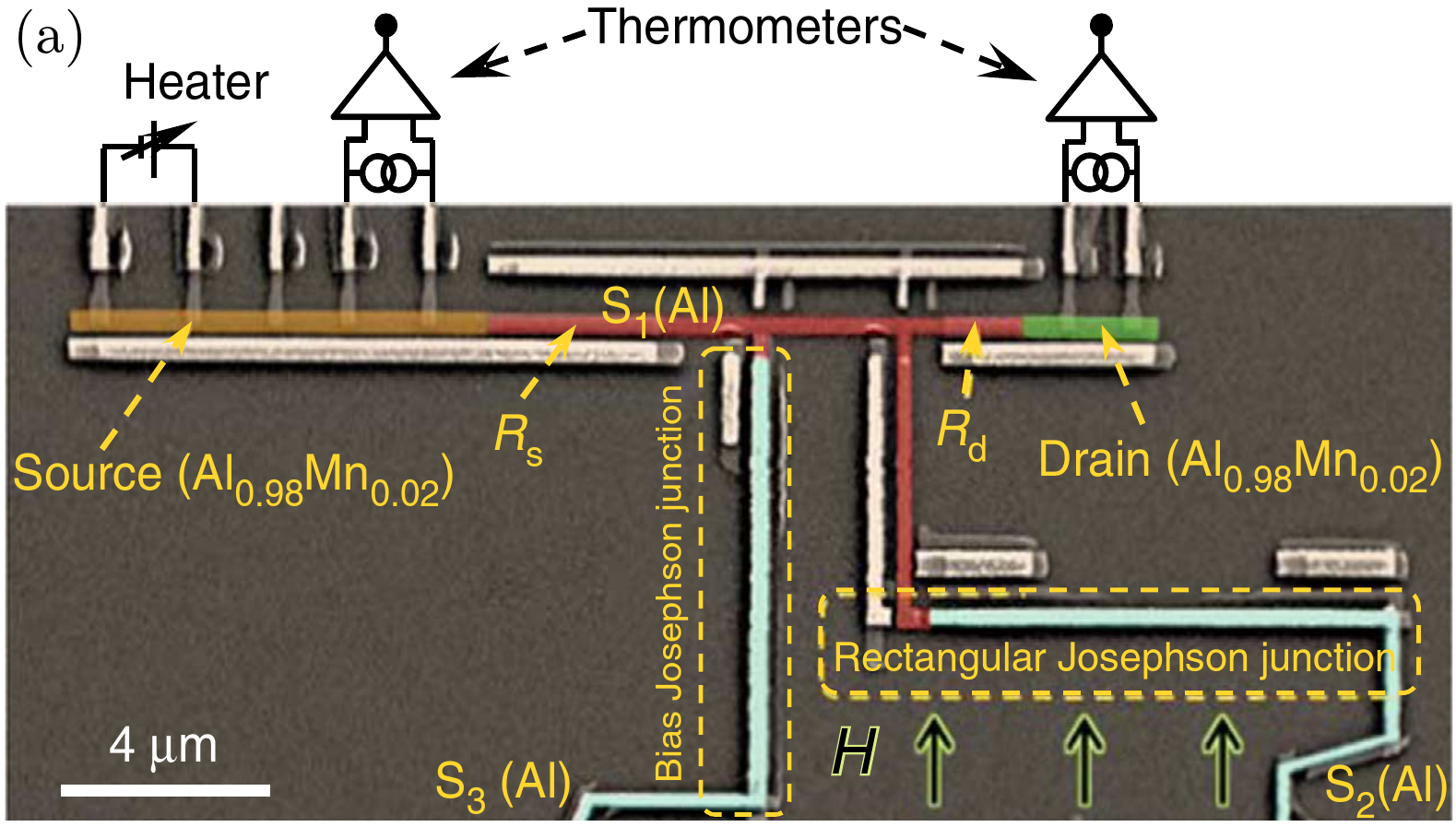}\\
\includegraphics[width=0.6\textwidth]{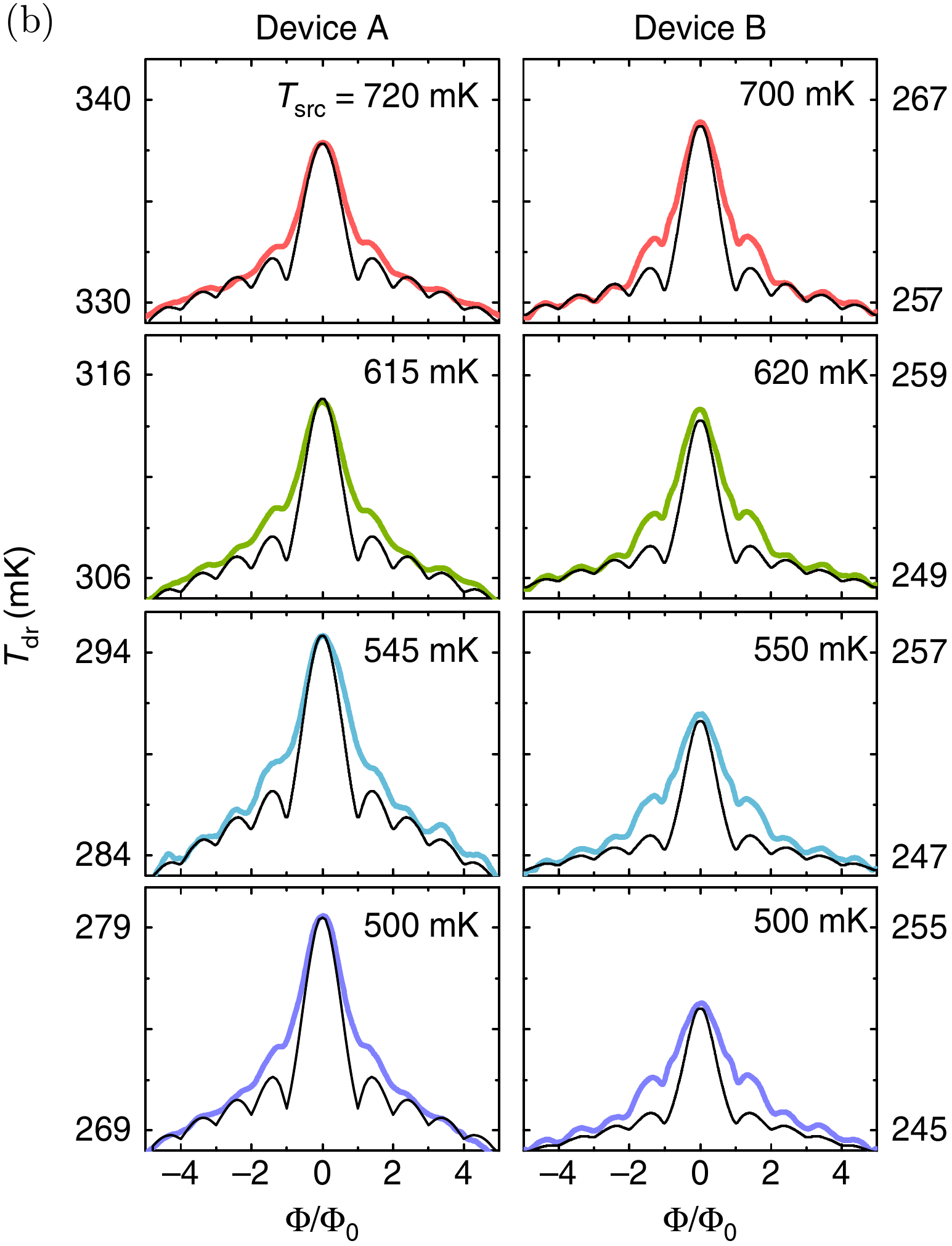}
\end{tabular}
\caption{ (a) A pseudo color image of the thermal diffractor with a rectangular Josephson junction, to which an in-plane magnetic field $H$ is applied. (b) Measured temperature diffraction patterns in the drain (colored lines) and the theoretical curves from the thermal model (black lines). Adapted from Ref.~\cite{martinez-perez_quantum_2014}.
}
\label{fig:ncom}
\end{figure}
In order to achieve good control over the phase bias across the Josephson junctions it is crucial that the two junctions forming the SQUID have identical resistances. Due to impurities in the fabrication process, however, even nominally identical junctions tend to have slightly different resistances. In order to overcome this issue and achieve a stronger modulation of heat currents with the magnetic flux, in a subsequent experiment a double SQUID consisting of a SQUID embedded into one of the arms of a SQUID has been used~\cite{fornieri_nanoscale_2016}. This allowed for a temperature modulation of about $\unit[40]{mK}$ and an almost ideal contrast in the phase modulation of heat currents with $(\dot Q_\text{AR}^\text{max}-\dot Q_\text{AR}^\text{min})/\dot Q_\text{AR}^\text{max}\simeq 99\%$. Remarkably, the flux-to-temperature transfer coefficient defined by $\partial T_\text{drain}/\partial\Phi$ can be as large as $\sim\unit[200]{mK}/\Phi_0$ which is much larger than previously reported values~\cite{giazotto_josephson_2012,martinez-perez_quantum_2014}.

While in the above experiments a good control over phase-coherent heat currents has been achieved, it was not possible to tune the phase-bias across a Josephson junction over the full range from $0$ to $\pi$. The latter would allow one to change the direction of the phase-depdendent heat current contribution from being parallel to the conventional quasiparticle contribution at $\phi=0$ to being antiparallel at $\phi=\pi$. A possibility to realize this full control over the phase was theoretically proposed in Ref.~\cite{fornieri_negative_2016}. The key idea is to use a superconducting ring interrupted by three Josephson junctions with critical currents $I_{c,i}$, $i=1,2,3$. If one of the junctions has a much smaller critical current than the other ones, say $I_{c,1}\ll I_{c,2}, I_{c,3}$, its phase difference can be tuned continuously from $0$ to $\pi$ by varying the magnetic flux through the ring by half a flux quantum. Experimentally, this proposal has been realized in Ref.~\cite{fornieri_0-pi_2017}. Remarkably, the additional control over the phase bias allowed for a relative temperature modulation $\delta T/T_\text{bath}$ of up to 400\% which is one order of magnitude larger than in the first experimental realization of a thermal interferometer~\cite{giazotto_josephson_2012}. Moreover, the device could be operated even at a bath temperature of $T_\text{bath}=\unit[800]{mK}$ where a temperature variation of $\delta T=\unit[20]{mK}$ has been observed.

The possibility to create a thermally biased Josephson junction with a phase difference that is tunable from $0$ to $\pi$ is not only important to increase the performance of thermal interferometers but it is also crucial to achieve a negative thermal differential conductance which is at the heart of realizing thermal transistors~\cite{fornieri_negative_2016}, thermal memories~\cite{guarcello_josephson_2018} and amplifiers~\cite{li_negative_2006,paolucci_phase-tunable_2017} and to ultimately build thermal logic gates~\cite{li_colloquium:_2012}.

\subsection{Josephson thermal diffractors}
When an extended Josephson junction is subject to a magnetic flux, the critical current of the junction will depend in a characteristic fashion on the flux. This Fraunhofer diffraction pattern can, e.g., be used to gather information about the spatial distribution of supercurrent flows~\cite{hart_induced_2014,pribiag_edge-mode_2015,amet_supercurrent_2016,bocquillon_gapless_2017}. A similar effect has been predicted for heat currents in extended Josephson junctions subject to a magnetic flux~\cite{giazotto_coherent_2013}. For example, for a rectangular junction, the flux dependence of the phase-dependent contribution to the heat current is given by the sine cardinal
\begin{equation}
	\dot Q(\Phi)=\dot Q_\text{max}\left|\frac{\sin \frac{\pi\Phi}{\Phi_0}}{\frac{\pi\Phi}{\Phi_0}}\right|.
\end{equation}
Thus, the heat flow $\dot Q(\Phi)$ exhibits minima whenever the applied flux $\Phi$ equals an integer multiple of the flux quantum $\Phi_0$.

The experimental detection of heat diffraction pattern was achieved with the setup shown in Fig.~\ref{fig:ncom}(a). It is similar to the original heat interferometer~\cite{giazotto_josephson_2012} in that it consists of two normal metal electrodes with superconducting electrodes for heating and temperature measurement. The central part of the setup is now given by an extended rectangular Josephson junction which is subject to an in-plane magnetic field. The resulting temperature modulation of the drain electrode is shown in Fig.~\ref{fig:ncom}(b) together with theoretically predicted temperatures based on a simple thermal model that accounts for electronic heat flows as well as electron-phonon couplings. The experimental findings are in good agreement with the theory and exhibit in particular the expected temperature minima at integer multiples of the flux quantum $\Phi_0$.
\begin{figure*}
\centering
\begin{minipage}{.7\textwidth}
\includegraphics[width=\textwidth]{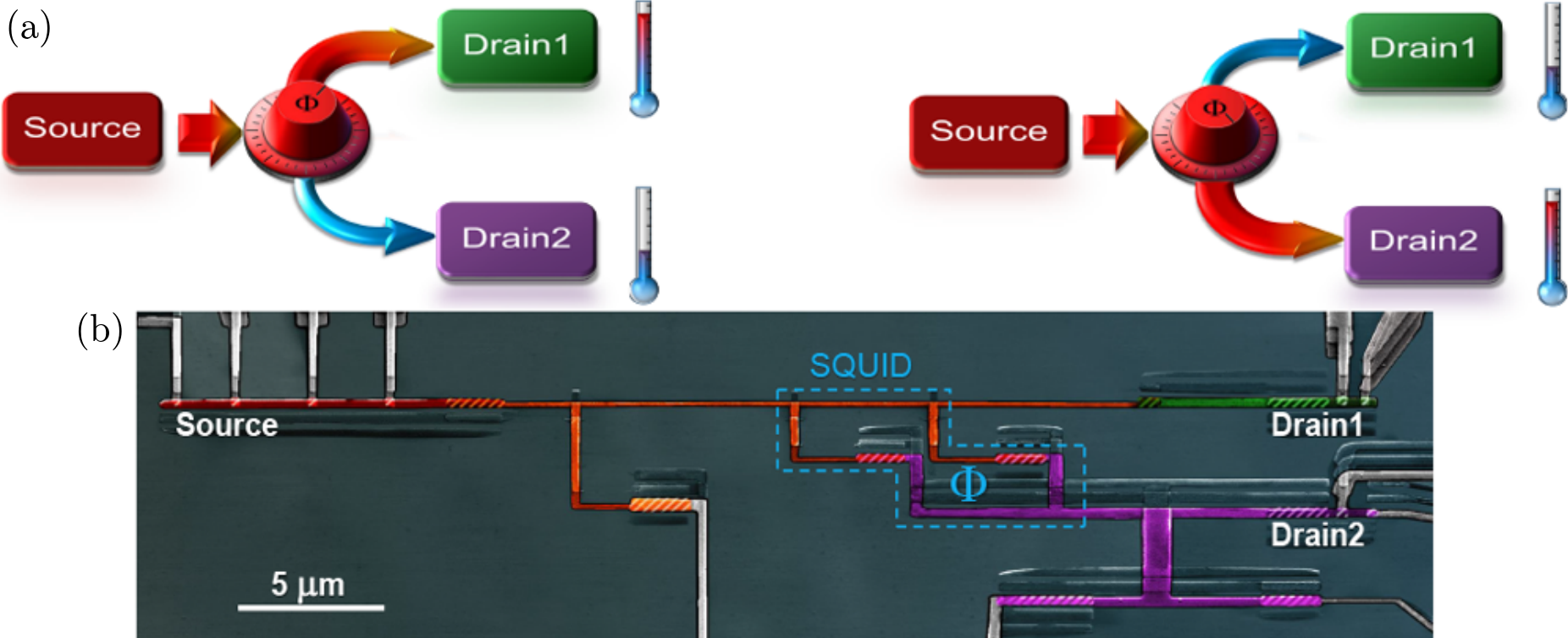}
\end{minipage}
\begin{minipage}{.29\textwidth}
\includegraphics[width=\textwidth]{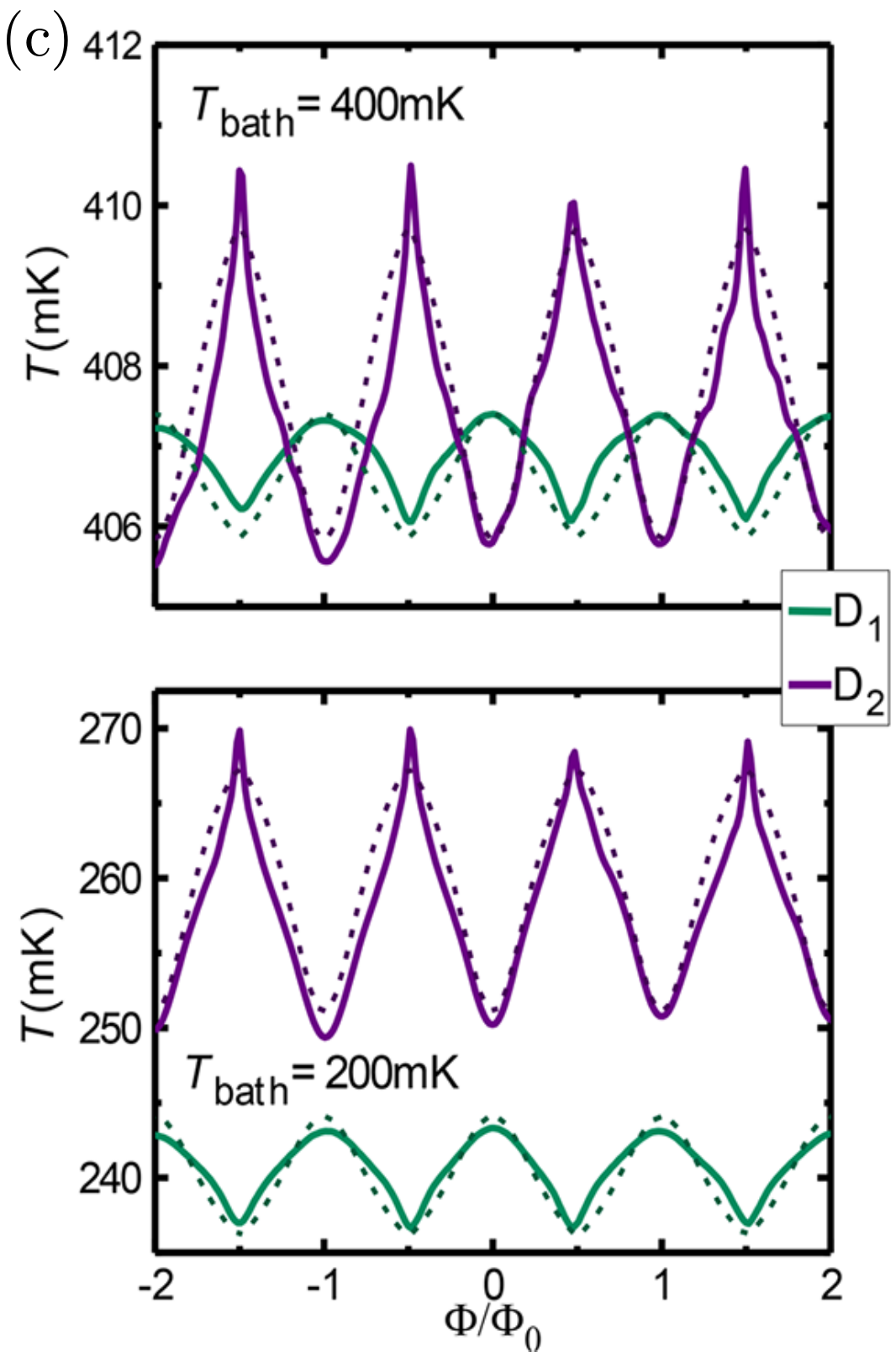}
\end{minipage}
\caption{(a) Schematics of a phase-tunable Josephson thermal router. With the magnetic flux $\Phi$ working as a knob, one can spatially distribute the source heat to the drain 1 or 2 on demand. (b) A false color micrograph image of the Josephson thermal router. (c) Measured temperature variation of the drain 1 (D1) and the drain 2 (D2) as a function of the magnetic flux at two different bath temperatures, in excellent agreement with the theoretical curves (dashed lines). Adapted from Ref.~\cite{timossi_phase-tunable_2018}.
}\label{fig:nanolett}
\end{figure*}
\subsection{Josephson thermal routers}
The possibility to control the magnitude of heat currents and, thus, to manipulate the temperature of superconducting terminals is an important feature of the various heat interferometers presented so far. An additional functionality that is of great importance for phase-coherent caloritronics is the operation of a phase-tunable thermal router which allows one to distribute an incoming heat current to different output terminals. Such a thermal router has been realized experimentally in Ref.~\cite{timossi_phase-tunable_2018}. The experimental setup shown in Fig.~\ref{fig:nanolett}(b) consists of a hot source connected to a first superconducting island $S_1$. The island is connected to a drain terminal but also forms part of a SQUID which involves a second superconducting island $S_2$ coupled to its respective drain. As shown schematically in Fig.~\ref{fig:nanolett}(a), heat from the source is directed towards drain 1 when the magnetic flux through the SQUID equals an integer multiple of the magnetic flux quantum, $\Phi=n\Phi_0$. In contrast, when the magnetic flux equals an half-integer multiple of the flux quantum, $\Phi=(n+1/2)\Phi_0$, the incoming heat is directed towards drain 2 instead.

The experimentally measured drain temperatures are shown in Fig.~\ref{fig:nanolett}(c) for two different base temperatures. Importantly, the device can operate in two different operation schemes depending on the base temperature. At low temperatures, one drain is always colder than the other. This regime is called the splitting regime. It appears for a bath temperature of \unit[200]{mK} and allows for a large temperature difference of about \unit[40]{mK} between the two drain electrodes. 
At higher bath temperatures, the device instead operates in the swapping regime. Here, at integer flux drain 1 is hotter than drain 2 while at half-integer flux drain 2 becomes hotter than drain 1. We remark that the temperature variation achievable in the swapping regime is much lower and reaches to only \unit[4]{mK}.

The thermal router is an important device for efficiently regulating the temperature of electrodes with a magnetic flux. The possibility to switch thermal signals among different channels can be an important ingredient in realizing thermal logic~\cite{paolucci_phase-tunable_2018}.

\section{Topological Josephson junctions}\label{STIS}
So far, we have reviewed the theoretical description of phase-dependent heat transport in S-N-S Josephson junctions as well as recent experimental progress on phase-coherent caloritroncs in such systems. We now turn to the discussion of phase-dependent thermal currents in topological Josephson junctions that have received a considerable amount of interest recently. In a topological Josephson junction, the central normal region consists of the surface state of a three-dimensional topological insulator or the one-dimensional helical edge states of a quantum spin Hall insulator, i.e. a two-dimensional topological insulator. 
Furthermore, it has recently become possible to form Josephson junctions where the central part is driven into the topologically nontrivial quantum Hall regime by a strong magnetic field~\cite{amet_supercurrent_2016,lee_inducing_2017,guiducci_toward_2019}.
In the following, we briefly summarize the differences in the theoretical description compared to the topologically trivial case and then report on how phase-dependent heat transport can be used to probe the presence of topologically nontrivial Andreev bound states~\cite{sothmann_fingerprint_2016} as well as on a thermal switch~\cite{sothmann_high-efficiency_2017} and rectifier~\cite{bours_phase-tunable_2019} based on topological Josephson junctions.

Topological Josephson junctions can be described by the Bogoliubov-de Gennes Hamiltonian $H_{\text{BdG}}$ in the basis of spin-up electrons, spin-down electrons, spin-up holes and spin-down holes analogous to Eq.~\eqref{eq:HBdG1} with
\beq\label{eq:HBdG2}
H_{\text{BdG}}=\begin{pmatrix}
h(x) & i\hat\sigma_y\Delta(x)\\
-i\hat\sigma_y\Delta^*(x) & -h^*(x)
\end{pmatrix},
\edq
where $\boldsymbol{\hat\sigma}=(\hat\sigma_x,\hat\sigma_y,\hat\sigma_z)$ is the vector of Pauli matrices and $\sigma_0$ denotes the identity matrix.
The single-particle Hamiltonian describing the topological surface states is given by 
\beq\label{eq:h2}
h(x)=-i\hbar v_F\hat\sigma_x\partial_x-\mu\hat\sigma_0,
\edq
where $v_F$ is the Fermi velocity. In contrast to the normal metal case in Eq.~\eqref{eq:h1}, the surface states of a topological insulator exhibit a linear dispersion relation. Furthermore, due to strong spin-orbit interactions, they are subject to spin-momentum locking, i.e., the spin orientation is locked to the propagation direction. This implies the absence of backscattering as long as time-reversal symmetry is preserved. 
The off-diagonal elements in Eq.~\eqref{eq:HBdG2} couple electrons and holes and arise from the bulk superconducting leads via the proximity effect.

\subsection{Signatures of topological Andreev bound states}
One of the most intriguing features of topological Josephson junctions is the formation of zero-energy Andreev bound states at a phase difference $\phi=\pi$~\cite{fu_superconducting_2008}. These zero-energy states are topologically protected, i.e., they are robust to the presence of disorder as long as time-reversal symmetry is preserved and can be described in terms of Majorana fermions that means as particles which are their own antiparticles. The topological Andreev bound states give rise to a  Josephson effect which is $4\pi$- rather than $2\pi$-periodic~\cite{fu_josephson_2009}. Experimentally, the detection of this fractional Josephson effect is challenging due to issues with quasiparticle poisoning and the presence of additional topologically trivial modes. Nevertheless, there has been a number of recent experiments that could indeed observe signatures of topologically nontrivial Andreev bound states~\cite{rokhinson_fractional_2012,sochnikov_nonsinusoidal_2015,wiedenmann_4pi-periodic_2016,bocquillon_gapless_2017,deacon_josephson_2017,li_4pi-periodic_2018,laroche_observation_2019,fornieri_evidence_2019,ren_topological_2019}

As an alternative way to probe the existence of such exotic low-energy Andreev bound states, in Ref.~\cite{sothmann_fingerprint_2016} it was suggested to investigate phase-dependent heat currents in topological Josephson junctions. In contrast to the fractional Josephson effect, heat transport measurements are immune to quasiparticle poisoning since heat currents are carried by quasiparticles with energies above the superconducting gap, i.e. the quasiparticle number is intrinsically fluctuating in a thermal transport scenario. Furthermore, it has been demonstrated in Ref.~\cite{sothmann_fingerprint_2016} that the characteristic signatures of zero-energy Andreev bound states are also clearly visible in the presence of additional gapped modes.
\begin{figure}
\centering
\centering
\begin{minipage}{.47\textwidth}
\includegraphics[width=\textwidth]{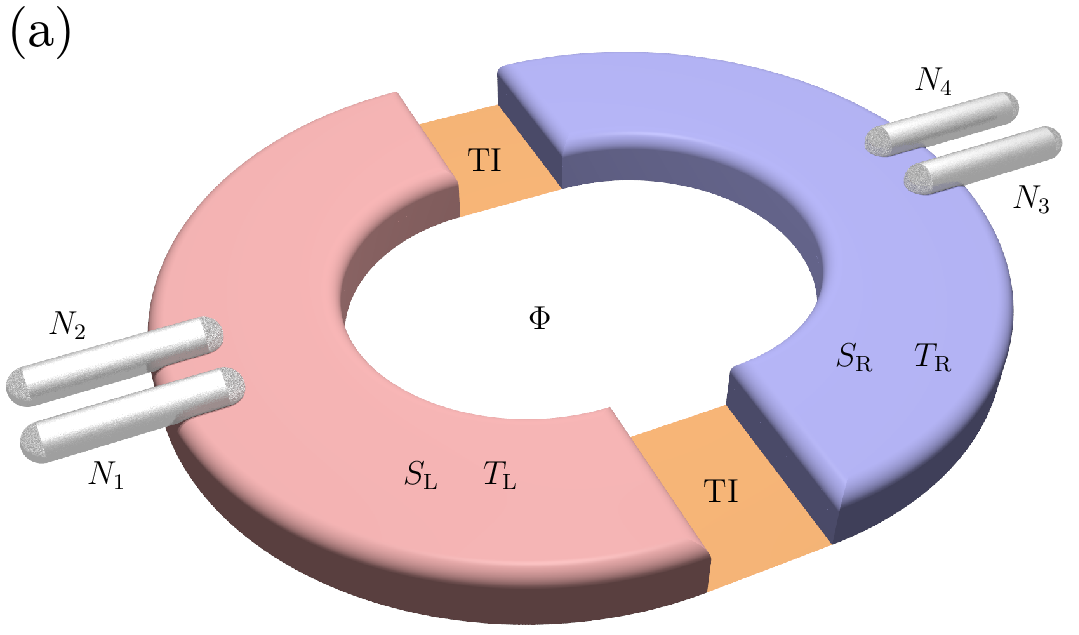}
\end{minipage}
\begin{minipage}{.5\textwidth}
\includegraphics[width=\textwidth]{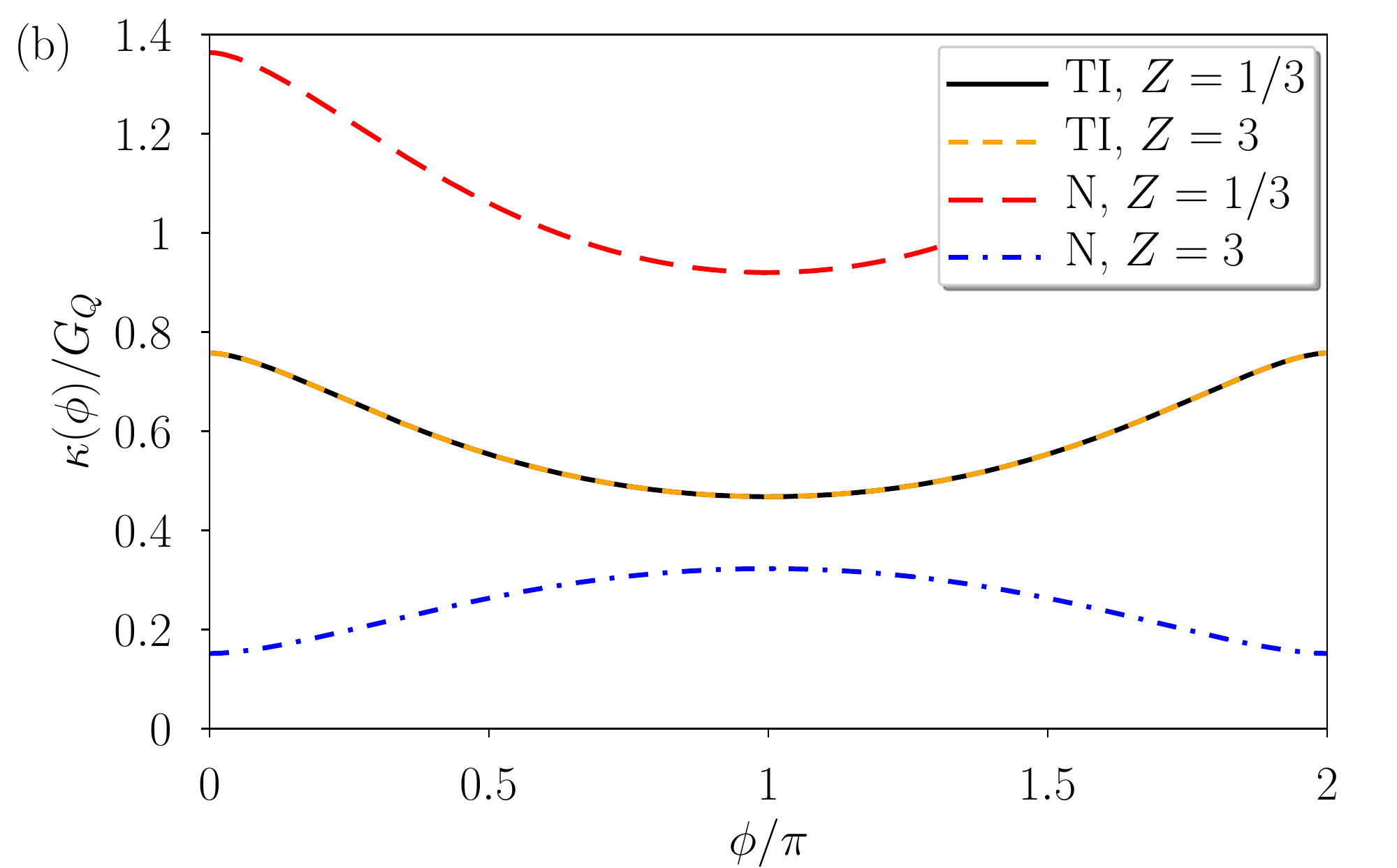}
\end{minipage}
%\begin{tabular}{cc}
%\includegraphics[width=0.45\textwidth]{Sothmann_PRBR_2016_1.png}&
%\includegraphics[width=0.5\textwidth]{Short_1D}
%\end{tabular}
\caption{(a) A setup with SQUID geometry to detect topological Andreev bound states via phase-dependent heat currents. The phase difference between two  superconductors $S_L$ and $S_R$ residing at respective temperatures $T_L$ and $T_R$ can be controlled by a magnetic flux $\Phi$. Normal metal electrodes $N_1$-$N_4$ act as heaters and thermometers. (b) Phase-dependent thermal conductance $\kappa(\phi)$ in units of thermal conductance quantum $G_Q$ at $k_BT=\Delta/2$ with different barrier transparencies $Z$ for a short one-dimensional S-N-S and S-TI-S junctions. Adapted from Ref. \cite{sothmann_fingerprint_2016}.
}
\label{fig:prb}
\end{figure}
The thermal conductance of a topological Josephson junction can be obtained in full analogy to the  trivial S-N-S case presented in Sec.~\ref{SNS} by solving the Bogoliubov-de Gennes equation based on the Hamiltonian in Eq.~\eqref{eq:HBdG2}. Matching the wave functions at the interface between the two superconductors gives rise to the transmission probabilities for electronlike and holelike quasiparticles, $\mathcal T_e(\omega,\phi)$ and $\mathcal T_h(\omega,\phi)$, respectively. They are given by
\beq\label{eq:Teh}
\calt_{e,h}(\omega,\phi)=\frac{\omega^2-\Delta^2}{\omega^2-\Delta^2\cos^2\frac{\phi}{2}}.
\edq
Importantly, the quasiparticle transmission probability in Eq. \eqref{eq:Teh} is independent of the interface potential barrier. This is in direct contrast to the transmission function of an S-N-S junction, cf. Eq.~\eqref{eq:TSNS}. The physical mechanism behind this is the spin-momentum locking caused by the strong spin-orbit coupling in the topological insulator. It forbids backscattering in the presence of time-reversal symmetry and, thus, leads to a superconducting analogue of Klein tunneling~\cite{tkachov_helical_2013} for quasiparticles. The transmission function in turn determines the phase-dependent thermal conductance
\beq
\kappa(\phi)=\frac{1}{h}\int d\omega\, \omega\sum_{i=e,h}\calt_i(\omega,\phi)\frac{df}{dT}.
\edq
It is shown in Fig.~\ref{fig:prb}(b) together with corresponding thermal conductances of S-N-S junctions. The thermal conductance of the S-TI-S junction is completely independent of the junction transmission $\tau=(1+Z^2)^{-1}$ in the normal state. In particular, it always exhibits a minimal value at phase difference $\pi$. This is in contrast to the behavior of a S-N-S junction where the thermal conductance changes from a minimum at $\phi=\pi$ in the transparent case to a maximum at $\phi=\pi$ in the tunneling limit. The reason for the robust minimum in the topological case is the existence of a protected zero-energy Andreev bound state. As detailed in Sec.~\ref{SNS}, it strongly reduces the transmission of quasiparticles above the superconducting gap and, thus, also the thermal conductance.

In the limit of low temperatures, $x=\Delta/k_BT\gg 1$, the minimal and maximal thermal conductance of a topological Josephson junction can be obtained analytically as
\begin{subequations}
	\begin{align}
		\kappa(\phi=0)&=\frac{2k_B^2T}{h}(x^2+2x+2)e^{-x},\\
        \kappa(\phi=\pi)&=\frac{4k_B^2T}{h}(x+1)e^{-x}.
	\end{align}
\end{subequations}
Both thermal conductances are exponentially suppressed due to the exponential suppression of quasiparticles available for thermal transport. While the maximal thermal conductance goes as $x^2$, the minimum thermal conductance goes as $x$. This means that by lowering the temperature, the amplitude of the conductance oscillation can be increased. As this has to compete against the overall exponential suppression of conductances at low temperatures, the optimal conditions for detecting the phase-dependence arise for intermediate temperatures $k_BT\sim \Delta/2$.

While the above discussion has been carried out for the case of a one-dimensional junction, it has been shown in Ref.~\cite{sothmann_fingerprint_2016} that qualitatively similar results also hold for a two-dimensional junction based on surface states of a three-dimensional topological insulator. This is again in contrast to S-N-S junction. Therefore, using the thermal conductance measurements, one can experimentally distinguish topological and trivial modes by adjusting the transparency of the junction as indicated in the schematic setup shown in Fig.~\ref{fig:prb}(a). Using realisitic parameters for this type of setup, it was estimated that a temperature variation of the cold superconducting electrode between \unit[360]{mK} and \unit[380]{mK} can be achieved for single-channel Josephson junctions which is well within the reach of present experimental measurement sensitivity.

% For a long junction with a finite junction length $L$ [cf. Eq. \eqref{eq:Teh}], distinct $\phi$-junction behaviors with completely different interference patterns provide clear fingerprints of topological Andreev bound states, as discussed in Ref.~\cite{sothmann_fingerprint_2016}.

\subsection{High-efficiency thermal switch}
\begin{figure}
\centering
\begin{tabular}{c}
\includegraphics[width=0.6\textwidth,clip]{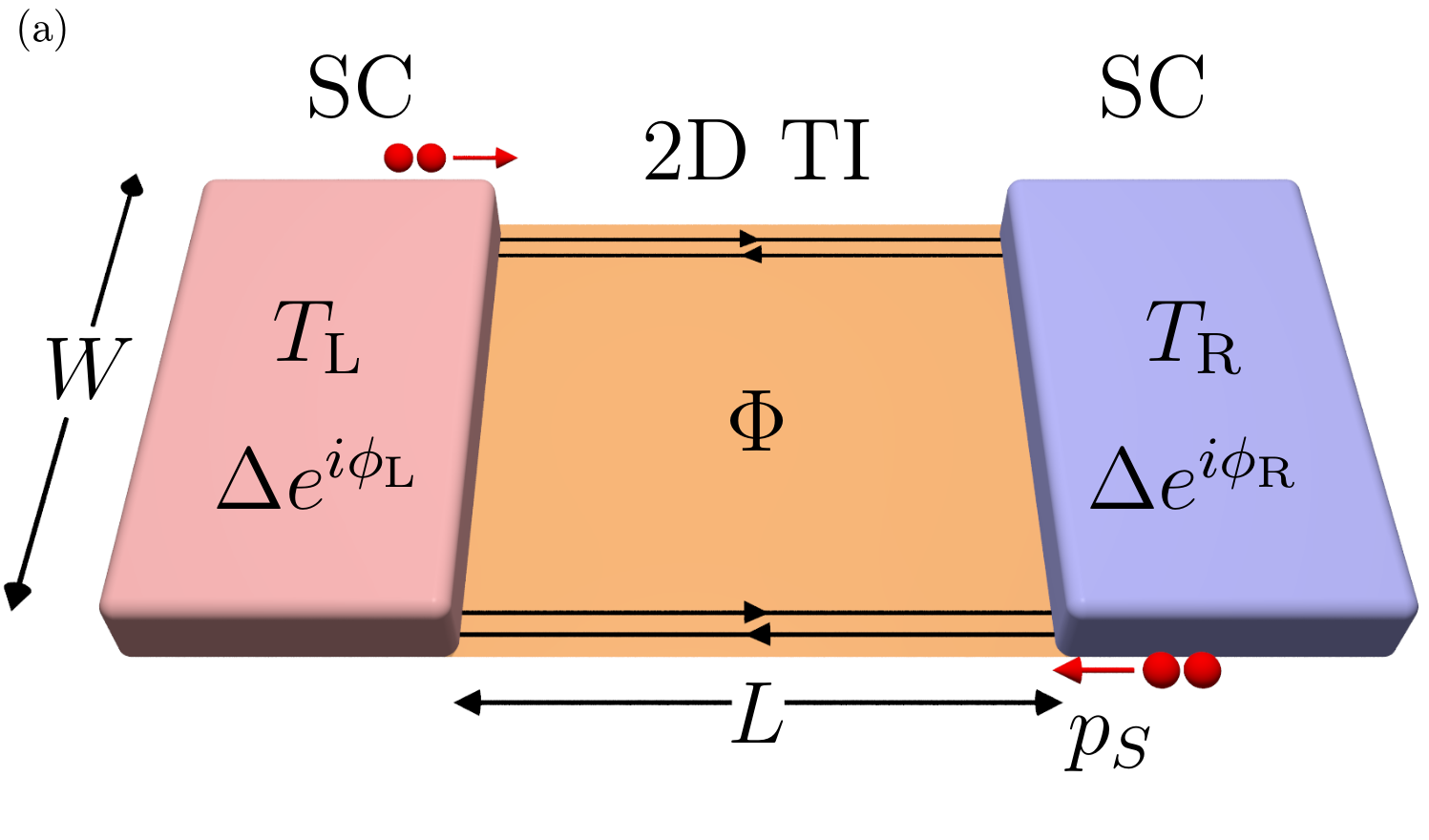}\\
\includegraphics[width=0.6\textwidth,clip]{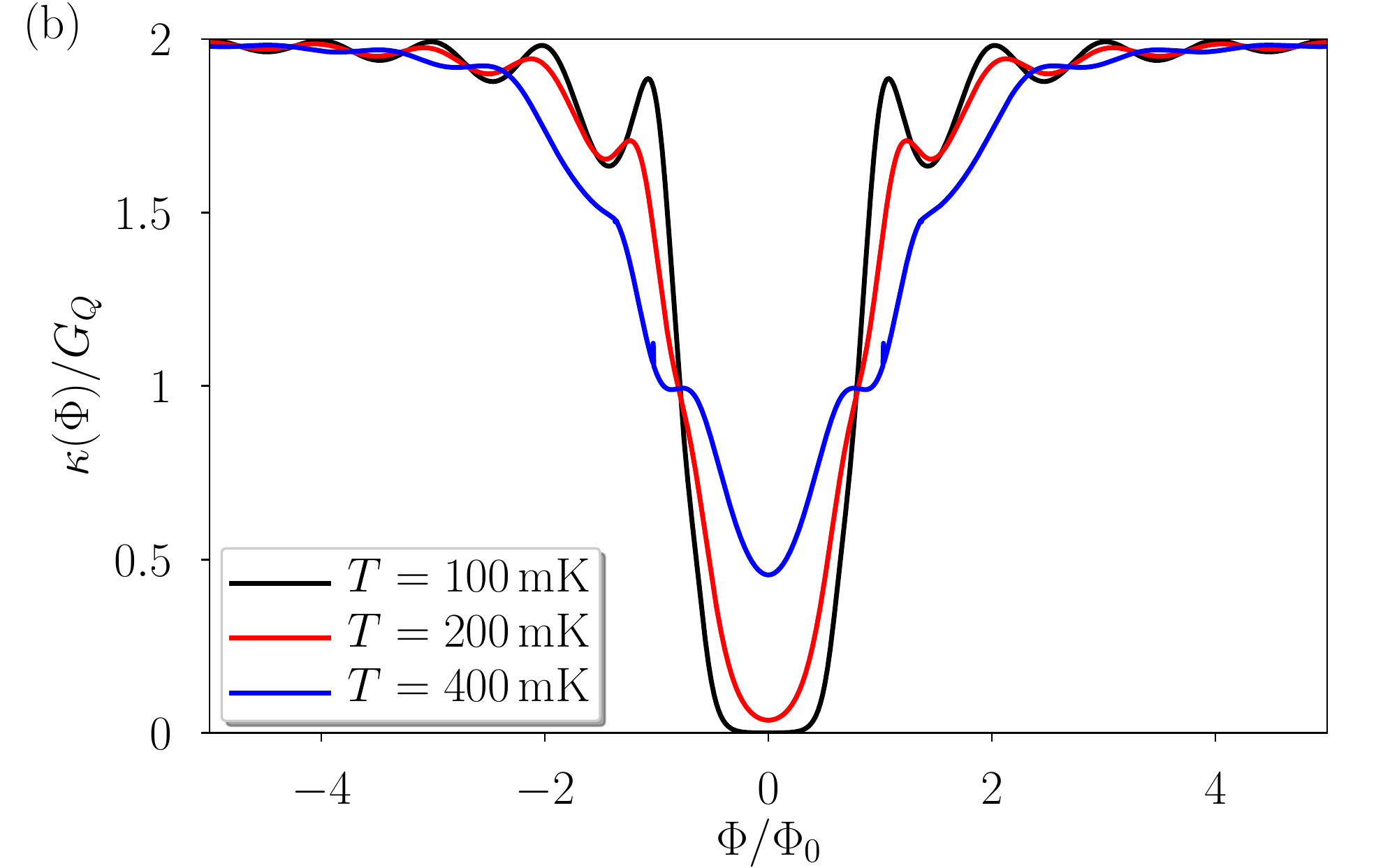}\\
\includegraphics[width=0.6\textwidth,clip]{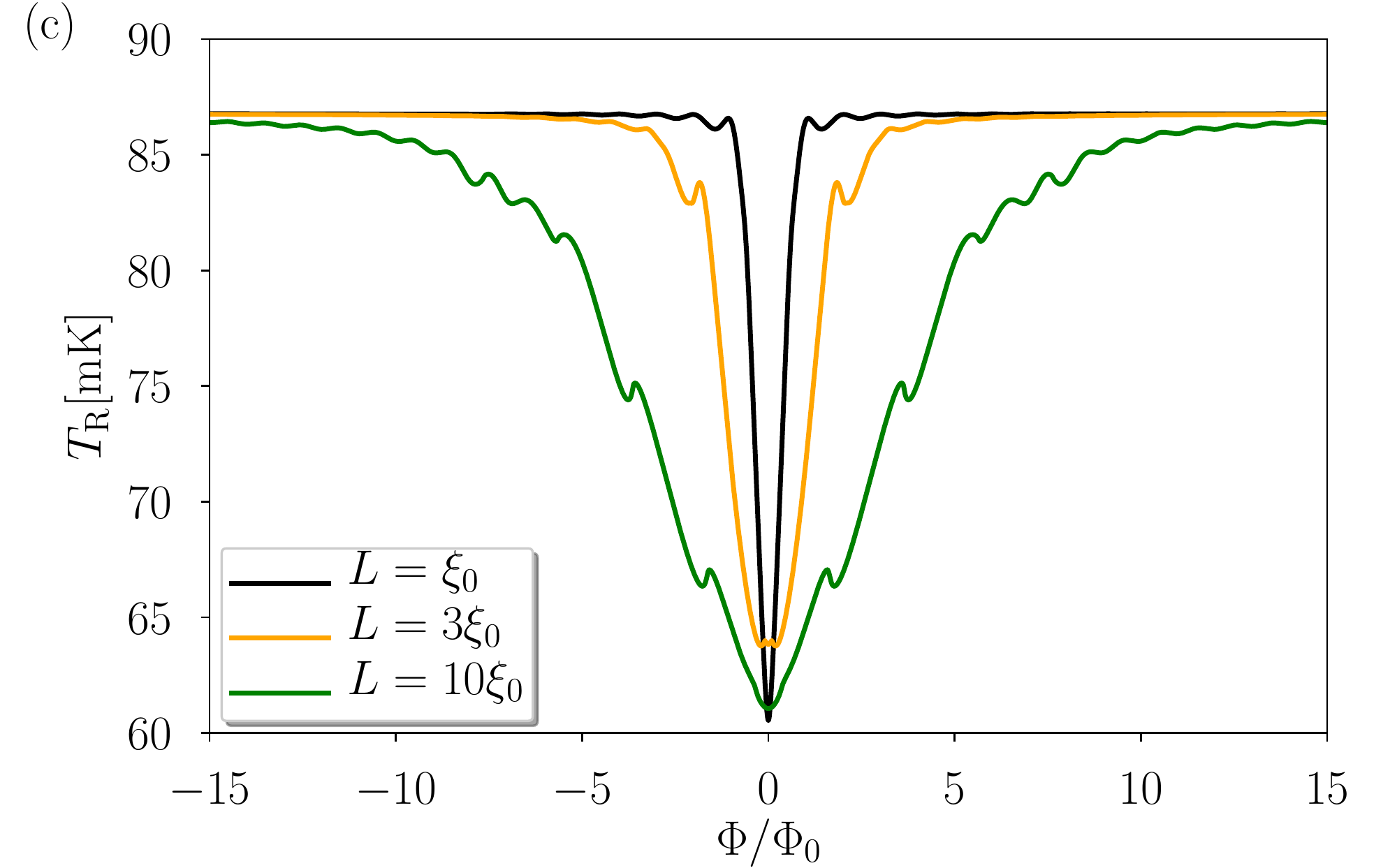}
\end{tabular}
\caption{(a) A Josephson junction based on a two-dimensional topological insulator. A magnetic flux $\Phi$ is applied to the junction of length $L$ and width $W$, which then gives rise to the Cooper pair momentum $p_S$ arising from screening currents inside the superconductors. (b) Flux-dependent thermal conductance $\kappa(\Phi)$ for a junction length $L={\xi }_{0}$ with several average temperatures $T$.
(c) Temperature variation of the right superconductor $T_\text{R}$ versus $\Phi$ for several junction lengths based on the simple thermal model with ${T}_{{\mathsf{bath}}}=50\,\mathrm{mK}$ and ${T}_{{\mathsf{L}}}=100\,\mathrm{mK}$. In (b) and (c), the results are shown with $\Delta=0.125$ meV and $\phi_0=0$.
Adapted from Ref.~\cite{sothmann_high-efficiency_2017}.
}
\label{fig:njp}
\end{figure}

In the previous section, we have reviewed how phase-coherent thermal transport can be used to gather information about fundamental physics such as the presence of Majorana fermions. At the same time, topological Josephson junctions can also be used to design devices for phase-coherent caloritroncs. For example, in Ref.~\cite{sothmann_high-efficiency_2017} it was shown that a topological Josephson junction subject to a magnetic flux can operate as a highly efficient thermal switch. The operational principle relies on the Doppler shift of the superconducting condensate which arises from screening currents that flow in response to the magnetic flux~\cite{tkachov_magnetic_2004,tkachov_quantum_2015}. The required magnetic fields are of the order of mT hence both, superconductivity and the helical edge conductance remain intact~\cite{pikulin_disorder_2014} with no backscattering induced~\cite{tkachov_ballistic_2010}. Importantly, the Dopper shift can close the induced superconducting gap in the edge states at a critical magnetic flux, thus lifting the exponentially suppressed thermal conductance of the superconducting state. Importantly, the gap closing occurs at one given critical flux for the edge channels which is in stark contrast to the case for systems with bulk modes where each mode has its own critical flux. Thus, the switching effect is robust with respect to the unintentional transport through the lowest bulk modes where each conducting channel closes at a different magnetic flux.

The corresponding setup is shown in Fig.~\ref{fig:njp}(a) and consists of a Josephson junction based on edge channels of a two-dimensional topological insulator. The junction has length $L$ and width $W$. The latter is assumed to large enough to guarantee a negligible overlap between helical edge channels on opposite sides of the junction. A magnetic field $B$ is applied perpendicular to the junction and gives rise to the threaded magnetic flux $\Phi=BLW$. 

The Hamiltonian describing the helical edge channels is modified in the presence of a magnetic flux and reads for the upper and lower helical edge states
\beq
h(x)=v_F\hat\sigma_x\left(-i\hbar \partial_x\pm\frac{p_S}{2}\right)-\mu\hat\sigma_0,
\edq
where
\beq
p_S=\frac{\pi\xi_0\Delta}{v_FL}\frac{\Phi}{\Phi_0}
\edq
is the screening-current-driven Cooper pair momentum due to the finite magnetic flux $\Phi$. Here $\xi_0=\hbar v_F/\Delta$ is the superconducting coherence length. Due to this additional Cooper pair momentum, the energies of right and left movers are Doppler-shifted by $v_Fp_S$ relative to each other. Hence, the gap of the whole system closes if this shift of energy is larger than the superconducting gap, i.e., if $v_Fp_S>2\Delta$. Moreover, in the presence of magnetic flux, the phase difference $\phi$ is not a gauge-invariant quantity but depends on the coordinates. One can however neglect the screening effects arising from the magnetic field generated from the Josephson current by assuming that the junction size is much smaller than the Josephson penetration depth (a good approximation for generic nanoscale junctions), in which case
\beq
\phi=\frac{2\pi y}{W}\frac{\Phi}{\Phi_0}+\phi_0,
\edq
with $\phi_0$ being the phase difference at $y=0$ which is the middle line along the junction with the edge channels located at $y=+W/2$ and $y=-W/2$. By solving for the eigenfunction of the Bogoliubov-de Gennes Hamiltonian and matching wave functions at the interface, one obtains the transmission function of the junction and, thus, the flux-dependent thermal conductance.

Figure~\ref{fig:njp}(b) shows the thermal conductance of a junction with length $L=\xi_0$ for three different base temperatures as a function of magnetic flux. At \unit[100]{mK}, the thermal conductance is exponentially suppressed around $\Phi=0$ due to the superconducting gap. When a small magnetic flux of $\Phi=(2/\pi)\Phi_0$ is applied, the thermal conductance is sharply increased to a value of about $\kappa\approx 2G_Q$ where $G_Q=\pi^2k_B^2T/(3h)$ is the quantum of thermal conductance. This abrupt switching behavior with a magnetic flux can be traced back to the aforementioned Doppler shift which can completely close the superconducting gap leading to a substantial increase in thermal conductance. At elevated temperatures, the thermal conductance remains finite even at vanishing flux due to thermally excited quasiparticles. Furthermore, the switching behavior is smeared out thermally. Nevertheless, the flux-controlled thermal switch operation still remains effective.
A qualitatively similar behavior can be observed for longer junctions. The main quantitative difference is that the switching with magnetic flux becomes smoother as the junction length is increased which can be used to control the switching characteristics.

To demonstrate the experimental feasibility of the proposed setup, in Fig.~\ref{fig:njp}(c) the temperature of the right superconductor is shown as a function of magnetic flux assuming that the left superconductor is heated to a constant temperature. In order to calculate the temperature of the right superconductor, a simple thermal model has been used which accounts for electronic heat currents as well as electron-phonon couplings where parameters have been estimated from existing experiments on superconducting caloritroncs. As can be seen, a temperature variation of about \unit[25]{mK} can be achieved upon varying the magnetic flux. This corresponds to a relative temperature varation of 
$\calr=(T_R^\text{max}-T_R^\text{min})/T_R^\text{min}\sim$ 40\%. This value is significantly larger than $\calr\sim$ 20\% which has been achieved in heat modulators based on conventional superconducting tunnel junctions~\cite{fornieri_nanoscale_2016}.
Furthermore, the proposed heat switch is robust against a variation of parameters even in the presence of bulk modes and hence provides an important ingredient in the toolkit of nanoscale thermal logic and heat management~\cite{fornieri_towards_2017}.

\subsection{Thermal rectification in TSQUIPT}
\begin{figure}
\centering
\begin{tabular}{c}
\includegraphics[width=0.55\textwidth]{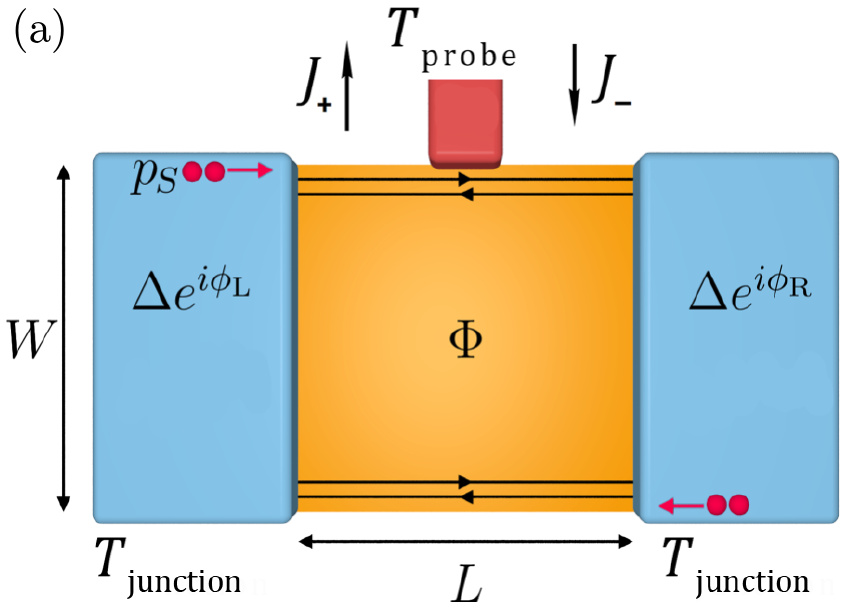}\\
\includegraphics[width=0.55\textwidth]{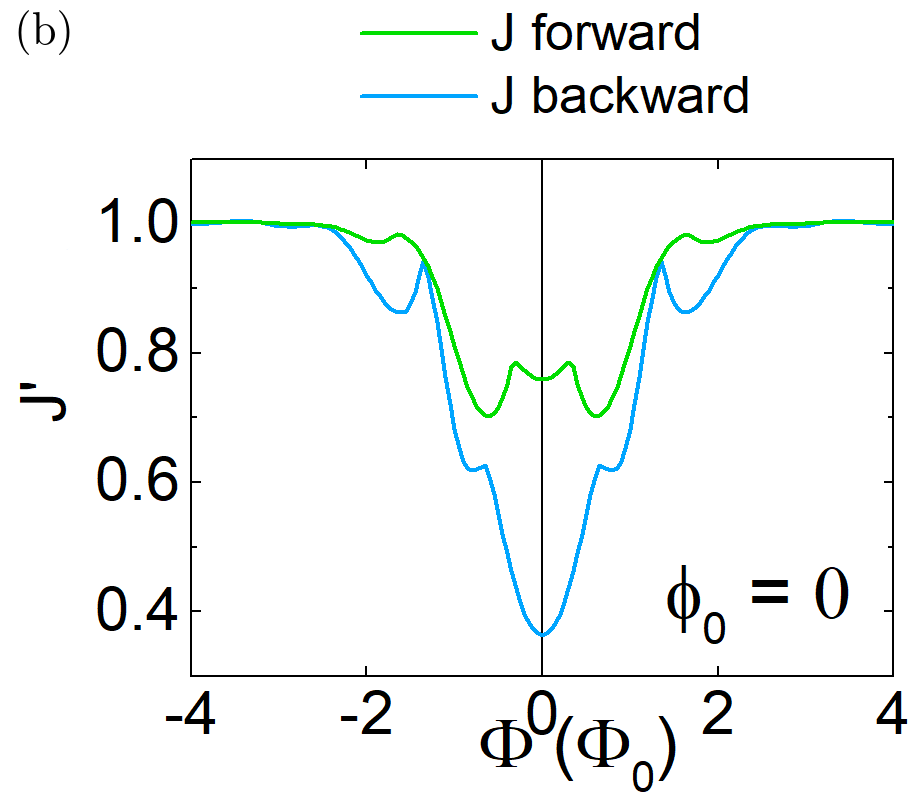}\\
\includegraphics[width=0.55\textwidth]{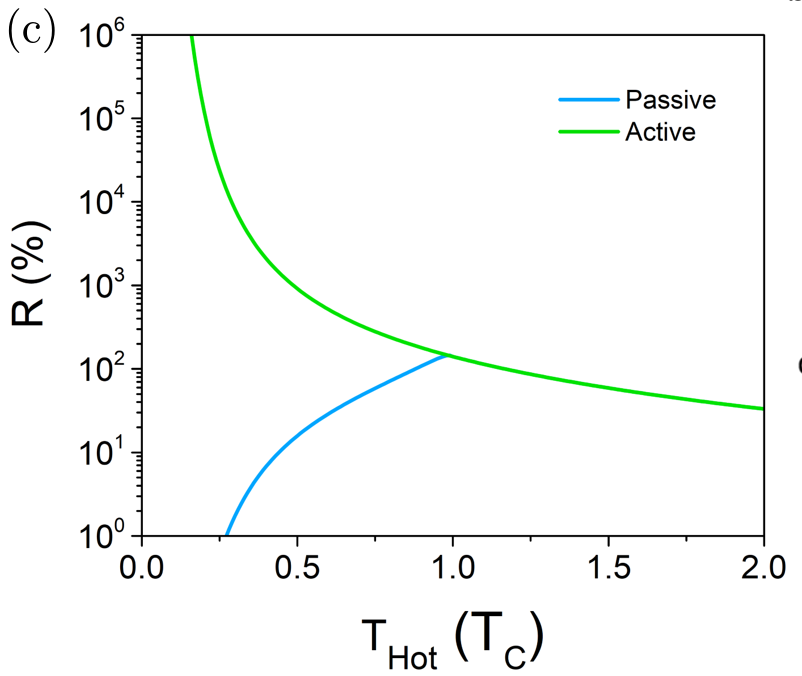}
\end{tabular}
\caption{(a) A sketch of TSQUIPT consisting of two superconductors on top of two-dimensional topological insulator to which a third normal metal probe is tunnel coupled. Two superconductors maintain the same temperature $T_\text{junction}$ while the probe with temperature $T_\text{probe}$ can be colder or hotter than the junction defining the forward ($J_+$) or backward ($J_-$) heat flow.
% $J_+$ and $J_-$ can then be asymmetric due to the temperature dependence of the proximity-induced superconducting gap on the helical edge states in contrast to the temperature-independent density of states of the probe terminal.
(b) The asymmetric forward and backward heat currents as a function of the magnetic flux $\Phi$ for a junction of length $L=\xi_0$ and $\phi_0=0$. In obtaining $J_+$, the hot and cold temperatures are respectively $T_\text{hot}=T_\text{junction}=0.9~T_C$ and $T_\text{cold}=T_\text{probe}=0.1~T_C$ where $T_C$ is the superconducting critical temperature. $J_-$ can be evaluated with a reversed temperature bias. (c) The relative rectification coefficient [Eq. \eqref{eq:rR}] with $L=\xi_0$, $\phi_0=0$, $T_\text{hot}=0.9~T_C$ and $T_\text{cold}=0.1~T_C$ for a passive mode ($\Phi=0$) and an active mode. For the latter, $\Phi=4\Phi_0$ for a forward bias whereas $\Phi=0$ for a backward bias configuration. Adapted from Ref.~\cite{bours_phase-tunable_2019}.
}
\label{fig:arxiv}
\end{figure}

A topological Josephson junction based on a two-dimensional topological insulator as introduced in the previous section can also be used to realize a so called topological superconducting quantum interference proximity transistor (TSQUIPT). It consists of an additional normal metal contact which is coupled to one of the edges of the topological insulator, cf. Fig.~\ref{fig:arxiv}(a). A TSQUIPT has been shown to provide a sensitive absolute magnetometer without the need for a ring structure~\cite{bours_topological_2018}. The latter is a great advantage over nontopological SQUIPTs~\cite{giazotto_superconducting_2010,strambini_omega-SQUIPT_2016,vischi_coherent_2017} and conventional SQUIDs in its implementation. The working principle of a TSQUIPT-based magnetometer is that the voltage drop generated across the system decays to a constant value with an applied magnetic flux in a nonperiodic fashion. This is in stark contrast to the $2\pi$-periodic flux dependence of magnetometers based on conventional SQUIDs which allows only for the measurement of relative field strength.

A thermal diode based on a TSQUIPT has been suggested recently~\cite{bours_phase-tunable_2019}. It can reach a rectification coefficient of up to 145\%. The rectification mechanism arises from the fact that the density of states of the edge channels aquires an implicit temperature dependence via the temperature dependence of the induced superconducting gap. At the same time, the density of states of the normal metal contact is temperature independent which gives rise to an asymmetric temperature response of the topological junction.

The full nonlinear heat current between the topological junction and the normal metal probe is given by
\beq
J=\frac{1}{e^2R_T}\int d\omega\omega\rho_\text{P}(\omega)\rho_\text{TI}(\omega)
	\left[f_\text{P}(\omega)-f_\text{TI}(\omega)\right],
\edq
where $\rho_\text{P}(\omega)$ and $\rho_\text{TI}(\omega)$ denote the density of states of the probe and the topological junction, respectively. Similarly, $f_\text{P}(\omega)$ and $f_\text{TI}(\omega)$ are the Fermi distribution of the probe and the superconducting leads. The above expression for the heat current is valid in the tunneling limit where the contact resistance satisfies $R_T\gg h/e^2$.

The normalized heat current $J'(\Phi)=J(\Phi)/J(\Phi\to\infty)$ flowing in response to a temperature bias between superconducting leads and the probe terminal is shown as a function of magnetic flux in Fig.~\ref{fig:arxiv}(b). It is asymmetric with respect to the direction of the temperature bias due to the temperature dependence of the superconducting gap. The relative rectification coefficient
\beq\label{eq:rR}
R(\%)=100\frac{|J_\text{+}|-|J_\text{-}|}{|J_\text{-}|}
\edq
can reach an optimal value of 145\%.
The rectification efficiency can be further increased by actively closing the quasiparticle gap only in the forward direction via the thermal switch effect proposed in Ref.~\cite{sothmann_high-efficiency_2017}. In this case, $R$ can reach values of up to $10^6\%$ for a temperature of the hot junction below the superconducting critical temperature as displayed in Fig. \ref{fig:arxiv}(c).

\subsection{Phase-coherent thermal circulators}
\begin{figure}
\centering
\includegraphics[width=0.5\textwidth,clip]{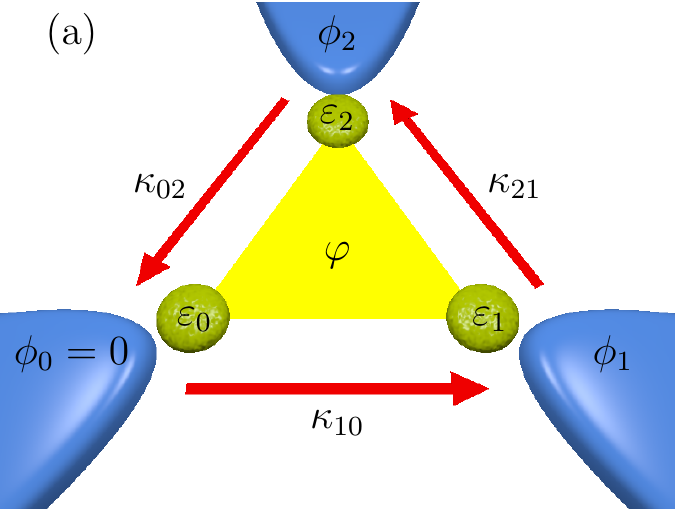}\\
\includegraphics[width=0.6\textwidth,clip]{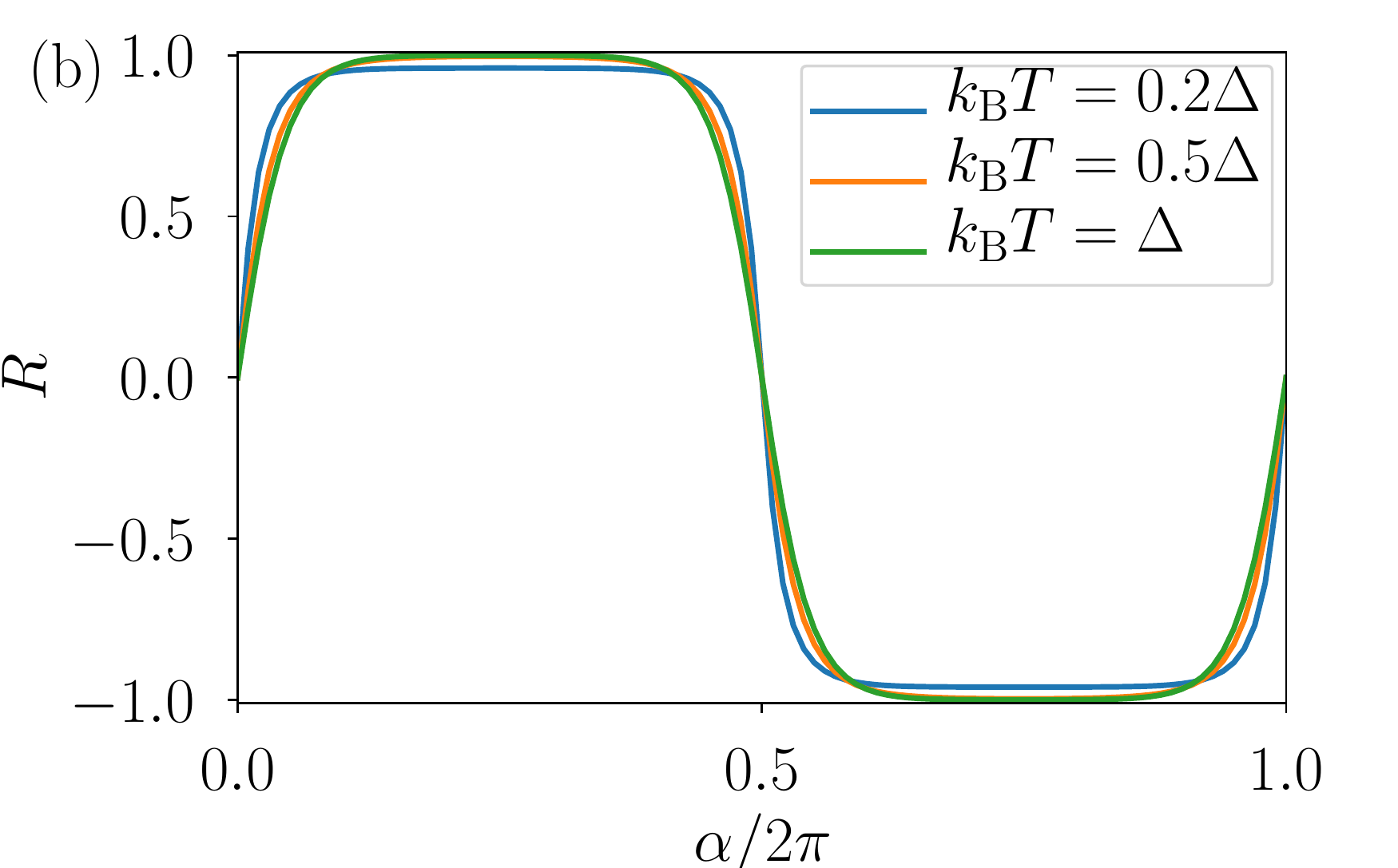}\\
\includegraphics[width=0.6\textwidth,clip]{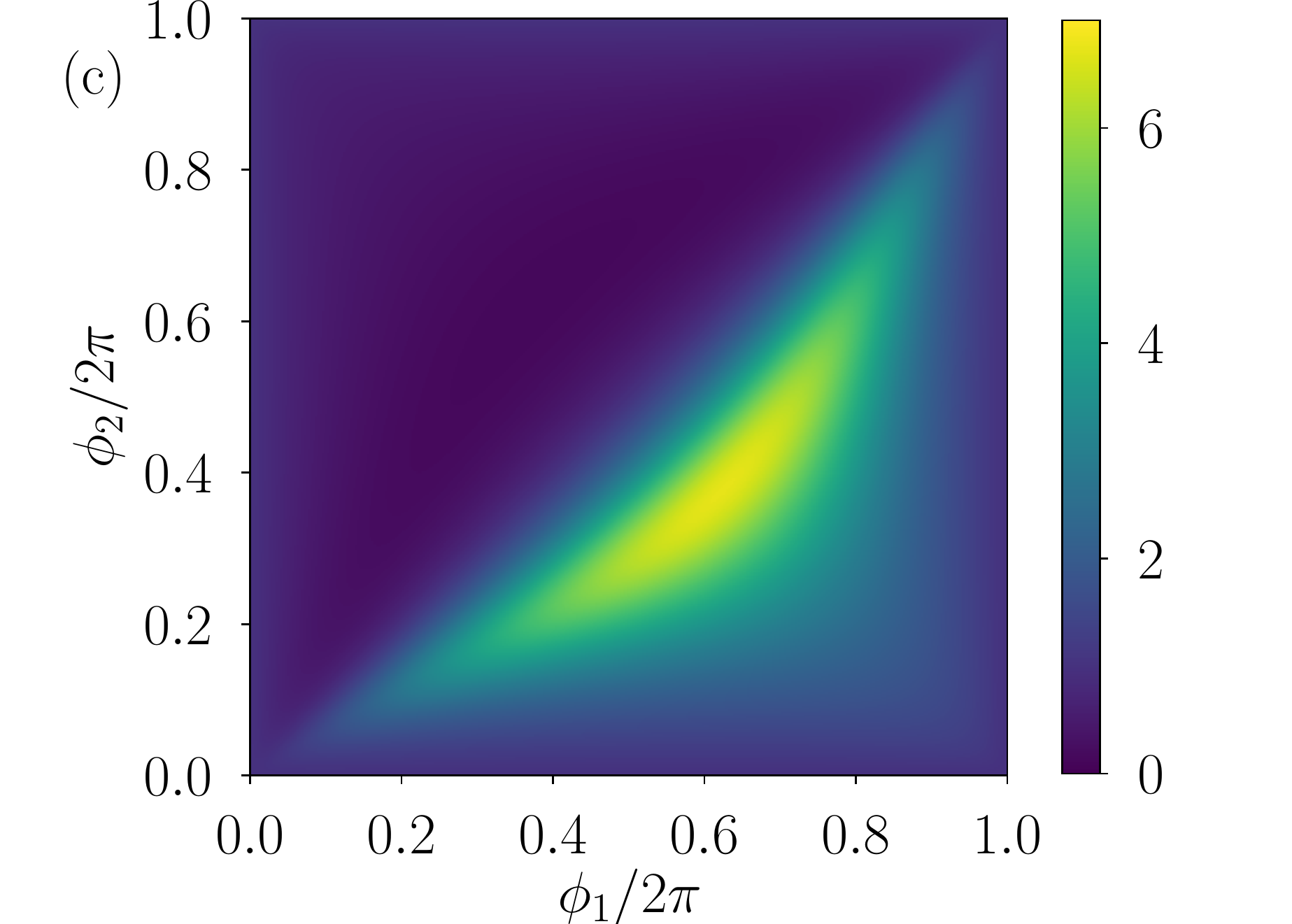}
\caption{(a) An illustration of the thermal circulator based on a three-terminal Josephson junction. (b) Rectification efficiency $R$ versus normalized magnetic flux $\alpha$ with various temperatures. (c) $\kappa_{21}/\kappa_{12}$ versus $\phi_1$ and $\phi_2$ at $k_BT=0.1\Delta$ and $\alpha=0$. Adapted from Ref.~\cite{hwang_phase-coherent_2018}.
}\label{fig:prappl}
\end{figure}

In addition to the growing interest in topological materials, there has been an interest in systems that allow for the simulation of topologically nontrivial band structures with trivial materials. A particular emphasis has been devoted to multi-terminal Josephson junctions~\cite{riwar_multi-terminal_2016,eriksson_topological_2017,meyer_nontrivial_2017,xie_topological_2017,xie_weyl_2018}. Here, the superconducting phase differences form the analogon of crystal momenta while the Andreev bound state energies correspond to the energy bands in a crystal. For a junction with at least four terminals, three independent superconducting phase differences form a sufficiently large number of degrees of freedom to mimic the behavior of Weyl points in the Andreev spectrum~\cite{riwar_multi-terminal_2016,eriksson_topological_2017,xie_weyl_2018}. Similary, in three-terminal junctions, the two superconducting phases together with a magnetic flux through the junction can be used to realize nontrivial Andreev bound state spectra of interest~\cite{meyer_nontrivial_2017,xie_topological_2017}.

In addition to the potential of simulating topological band structures, multi-terminal Josephson junctions are also of interest for phase-coherent caloritronics~\cite{hwang_phase-coherent_2018}. In particular, they allow for the realization of heat circulators which are analogous to microwave circulators in electronics~\cite{hogan_ferromagnetic_1953}. Compared to circulators based on the quantum Hall effect which requires large magnetic fields of several Tesla~\cite{viola_hall_2014,mahoney_-chip_2017}, systems based on multi-terminal Josephson junctions offer the advantage of relying on much smaller magnetic fields which are compatible with superconductivity and thus allow for the integration into other phase-coherent caloritronic devices.

A possible minimal setup for such a heat circulator is shown in Fig.~\ref{fig:prappl}(a). It consists of three sites with energies $\varepsilon_i$, $i=0,1,2$ which are tunnel coupled among each other. Each site is also coupled to a superconducting electrode whose order parameter has the phase $\phi_i$. Furthermore, the junction is subject to a perpendicular magnetic field that gives rise to the flux $\varphi$. The thermal conductance from terminal $j$ to terminal $i$ is given by
\beq
\kappa_{ij}(\phi_1,\phi_2)=\frac{1}{hT}\int d\omega~\omega^2\left[2\delta_{ij}-\calt_{ij}\right]
	\left(-\frac{\partial f}{\partial\omega}\right).
\edq
The transmission probability $\mathcal T_{ij}$ depends on energy and magnetic flux as well as on the superconducting phases $\phi_i$. Without loss of generality, we can set $\phi_0=0$ such that the thermal conductance is controlled by the two phases $\phi_1$ and $\phi_2$ independently. Importantly, in the presence of magnetic field $\bf B=\nabla\times\bf A$, one should consider the gauge-invariant phase $\gamma_{ij}=\phi_i-\phi_j-2\pi/(h/2e)\int_j^i{\bf A}\cdot d{\bf s}$ where the integration of the vector potential ${\bf A}$ can be taken along an arbitrary path $d{\bf s}$ connecting the two superconductors $i$ and $j$.

Having evaluated all the thermal conductances among the superconductors, i.e., $\kappa_{ij}$ ($i,j=0,1,2$), the efficiency of the circulator can be quantified by
\beq\label{eq:R}
R=\frac{\kappa_{\Circlearrowleft}-\kappa_{\Circlearrowright}}{\kappa_{\Circlearrowleft}+\kappa_{\Circlearrowright}},
\edq
where $\kappa_{\Circlearrowleft}=\kappa_{02}\kappa_{21}\kappa_{10}$ and $\kappa_{\Circlearrowright}=\kappa_{01}\kappa_{12}\kappa_{20}$ respectively describe the counterclockwise and clockwise heat circulation along the junction. Thus, $R$ vanishes if there is no heat circulation as $\kappa_{\Circlearrowleft}=\kappa_{\Circlearrowright}$ in Eq. \eqref{eq:R}, while $|R|=1$ for either a perfect clockwise ($R=-1$) or a perfect counterclockwise ($R=1$) heat circulation.

In Fig. \ref{fig:prappl}(b), $R$ is shown as a function of the normalized magnetic flux $\alpha=\varphi/\phi_0$ with $\phi_0=e/h$. Remarkably, the device exhibits almost an ideal efficiency with counterclockwise heat circulation, i.e., $R\approx 1$ upon application of a small magnetic flux $0<\alpha<\pi$. When half a magnetic flux quantum penetrates the junction, the circulation efficiency vanishes due to symmetry. When the magnetic flux is further increased $\pi<\alpha<2\pi$, the device shows again an almost perfect heat circulation which is now clockwise, $R\approx -1$. Hence, the direction of heat circulation can be manipulated easily without the need for reversing a large magnetic field. The thermal conductance rectification $\kappa_{21}/\kappa_{12}$ can actually be as large as 100 without breaking the superconductivity with an increased average temperature. This is mainly due to the thermally excited quasiparticles at elevated temperatures. The heat flow in the forward direction is estimated to be of the order of 100 fW with Al superconductors and a temperature gradient 100 mK.

Figure \ref{fig:prappl}(c) displays the rectification $\kappa_{21}/\kappa_{12}$, tuned with two superconducting phases $\phi_1$ and $\phi_2$ when no magnetic field is applied reaching values as large as 7 in an optimum phase bias. Thus, the time-reversal symmetry breaking by superconducting phases can also realize a good circulator.

In Ref.~\cite{hwang_phase-coherent_2018}, it is also shown that the circulator efficiency is quite robust with respect to the disorder. As this device has been shown to be a prototype for simulating topologically nontrivial band structures~\cite{meyer_nontrivial_2017,xie_topological_2017}, it will be worthwhile to explore the topological nature in thermal conductance of the multiterminal Josephson junctions.

\section{Conclusions}\label{sum}
In this topical review, we have discussed both theoretical and experimental developments of phase-dependent heat transport in ordinary and topological Josephson junctions.
Starting from the theoretical description of the phase dependence of the thermal conductance in the simplest example of a short, one-dimensional S-N-S junction with a Bogloiubov-de Gennes description, we have reviewed recent key experiments in ordinary Josephson junctions based on various dc SQUID structures. Finally, we have discussed theoretical developments in the phase-coherent caloritronics with topological Josephson junctions where the phase-dependent thermal conductance can be used not only to gather information about topological Andreev bound states but also to implement new device functionalities. 

Importantly, the half-a-century gap between the first theoretical prediction of phase-dependent heat transport in Josephson junctions and its experimental verification has been finally filled in recent years. This has proven to be the starting point of a new field in nanoscience, namely phase-coherent caloritonics. It can be expected that in the near future many other caloritronic devices will be proposed and, subsequently, realized experimentally. 
Understanding the influence of electron-electron interactions on phase-dependent heat transport in both topological and trivial Josephson junctions provides an interesting direction for future research.
Ultimately, nanoscale caloritronics can help in achieving an efficient management of waste heat in nanostructures and can, thus, help to contribute to future efficient electronic devices.

\begin{acknowledgement}
This work was supported by the Ministry of Innovation NRW via the ``Programm zur F\"orderung der R\"uckkehr des hochqualifizierten Forschungsnachwuchses aus dem Ausland''.
\end{acknowledgement}
% The section below may be edited at your convenience to acknowledge 
% each author's contribution to the manuscript.
% You may remove it if you are a single author.
%
%\section{Authors contributions}
%All the authors were involved in the preparation of the manuscript.
%All the authors have read and approved the final manuscript.
%
\section*{Author contribution statement}
Both authors contributed equally to this review article.

% BibTeX users please use
% \bibliographystyle{epj}
% \bibliography{/home/bjoern/LaTeX/Bibtex/Meine_Bibliothek.bib}
% \bibliography{Meine_Bibliothek}
%\bibliography{epjreview}

\begin{thebibliography}{120}

\bibitem{fourier_theorie_1822}
J.B.J.b. Fourier, \emph{Théorie analytique de la chaleur} (Chez Firmin Didot,
  père et fils, 1822)

\bibitem{seebeck_ueber_1826}
T.J. Seebeck, Ann. Phys. \textbf{82}, 253 (1826)

\bibitem{hicks_effect_1993}
L.D. Hicks, M.S. Dresselhaus, Phys. Rev. B \textbf{47}, 12727 (1993)

\bibitem{hicks_thermoelectric_1993}
L.D. Hicks, M.S. Dresselhaus, Phys. Rev. B \textbf{47}, 16631 (1993)

\bibitem{mahan_best_1996}
G.D. Mahan, J.O. Sofo, Proc. Natl. Acad. Sci. USA \textbf{93}, 7436 (1996)

\bibitem{staring_coulomb-blockade_1993}
A.A.M. Staring, L.W. Molenkamp, B.W. Alphenaar, H.~van Houten, O.J.A. Buyk,
  M.A.A. Mabesoone, C.W.J. Beenakker, C.T. Foxon, Europhysics Letters (EPL)
  \textbf{22}, 57 (1993)

\bibitem{venkatasubramanian_thin-film_2001}
R.~Venkatasubramanian, E.~Siivola, T.~Colpitts, B.~O'Quinn, Nature
  \textbf{413}, 597 (2001)

\bibitem{schwab_measurement_2000}
K.~Schwab, E.A. Henriksen, J.M. Worlock, M.L. Roukes, Nature \textbf{404}, 974
  (2000)

\bibitem{meschke_single-mode_2006}
M.~Meschke, W.~Guichard, J.P. Pekola, Nature \textbf{444}, 187 (2006)

\bibitem{jezouin_quantum_2013}
S.~Jezouin, F.D. Parmentier, A.~Anthore, U.~Gennser, A.~Cavanna, Y.~Jin,
  F.~Pierre, Science \textbf{342}, 601 (2013)

\bibitem{dutta_thermal_2017}
B.~Dutta, J.~Peltonen, D.~Antonenko, M.~Meschke, M.~Skvortsov, B.~Kubala,
  J.~König, C.~Winkelmann, H.~Courtois, J.~Pekola, Phys. Rev. Lett.
  \textbf{119}, 077701 (2017)

\bibitem{ronzani_tunable_2018}
A.~Ronzani, B.~Karimi, J.~Senior, Y.C. Chang, J.T. Peltonen, C.~Chen, J.P.
  Pekola, Nat. Phys. p.~1 (2018)

\bibitem{wang_fast_2018}
L.B. Wang, O.P. Saira, J.P. Pekola, Appl. Phys. Lett. \textbf{112}, 013105
  (2018)

\bibitem{maillet_optimal_2019}
O.~Maillet, P.A. Erdman, V.~Cavina, B.~Bhandari, E.T. Mannila, J.T. Peltonen,
  A.~Mari, F.~Taddei, C.~Jarzynski, V.~Giovannetti et~al., Phys. Rev. Lett.
  \textbf{122}, 150604 (2019)

\bibitem{shakouri_recent_2011}
A.~Shakouri, Annu. Rev. Mater. Res. \textbf{41}, 399 (2011)

\bibitem{radousky_energy_2012}
H.B. Radousky, H.~Liang, Nanotechnology \textbf{23}, 502001 (2012)

\bibitem{kosloff_quantum_2014}
R.~Kosloff, A.~Levy, Annu. Rev. Phys. Chem. \textbf{65}, 365 (2014)

\bibitem{sanchez_focus_2014}
D.~Sánchez, H.~Linke, New J. Phys. \textbf{16}, 110201 (2014)

\bibitem{sothmann_thermoelectric_2015}
B.~Sothmann, R.~Sánchez, A.N. Jordan, Nanotechnology \textbf{26}, 032001
  (2015)

\bibitem{pekola_towards_2015}
J.P. Pekola, Nat. Phys. \textbf{11}, 118 (2015)

\bibitem{sanchez_nonlinear_2016}
D.~Sánchez, R.~López, C. R. Phys. \textbf{17}, 1060 (2016)

\bibitem{benenti_fundamental_2017}
G.~Benenti, G.~Casati, K.~Saito, R.~Whitney, Phys. Rep. \textbf{694}, 1 (2017)

\bibitem{pekola_thermodynamics_2019}
J.~Pekola, I.~Khaymovich, Annu. Rev. Cond. Mat. Phys. \textbf{10}, 193 (2019)

\bibitem{rowe_CRC_1995}
D.M. Rowe, ed., \emph{{CRC} {Handbook} of {Thermoelectrics}}, 1st~edn. (CRC
  Press, Boca Raton, etc., 1995), ISBN 978-0-8493-0146-9

\bibitem{beekman_better_2015}
M.~Beekman, D.T. Morelli, G.S. Nolas, Nat. Mater. \textbf{14}, 1182 (2015)

\bibitem{bourgeois_reduction_2016}
O.~Bourgeois, D.~Tainoff, A.~Tavakoli, Y.~Liu, C.~Blanc, M.~Boukhari,
  A.~Barski, E.~Hadji, C. R. Phys. \textbf{17}, 1154 (2016)

\bibitem{fornieri_towards_2017}
A.~Fornieri, F.~Giazotto, Nature Nanotech. \textbf{12}, 944 (2017)

\bibitem{josephson_possible_1962}
B.D. Josephson, Phys. Lett. \textbf{1}, 251 (1962)

\bibitem{maki_entropy_1965}
K.~Maki, A.~Griffin, Phys. Rev. Lett. \textbf{15}, 921 (1965)

\bibitem{maki_entropy_1966}
K.~Maki, A.~Griffin, Phys. Rev. Lett. \textbf{16}, 258 (1966)

\bibitem{guttman_phase-dependent_1997}
G.D. Guttman, B.~Nathanson, E.~Ben-Jacob, D.J. Bergman, Phys. Rev. B
  \textbf{55}, 3849 (1997)

\bibitem{guttman_interference_1998}
G.D. Guttman, E.~Ben-Jacob, D.J. Bergman, Phys. Rev. B \textbf{57}, 2717 (1998)

\bibitem{golubev_heat_2013}
D.~Golubev, T.~Faivre, J.P. Pekola, Phys. Rev. B \textbf{87}, 094522 (2013)

\bibitem{zhao_phase_2003}
E.~Zhao, T.~Löfwander, J.A. Sauls, Phys. Rev. Lett. \textbf{91}, 077003 (2003)

\bibitem{zhao_heat_2004}
E.~Zhao, T.~Löfwander, J.A. Sauls, Phys. Rev. B \textbf{69}, 134503 (2004)

\bibitem{giazotto_phase-tunable_2013}
F.~Giazotto, F.S. Bergeret, Appl. Phys. Lett. \textbf{102}, 132603 (2013)

\bibitem{bergeret_phase-dependent_2013}
F.S. Bergeret, F.~Giazotto, Phys. Rev. B \textbf{88}, 014515 (2013)

\bibitem{virtanen_thermal_2014}
P.~Virtanen, F.~Giazotto, Phys. Rev. B \textbf{90}, 014511 (2014)

\bibitem{virtanen_fluctuation_2015}
P.~Virtanen, F.~Giazotto, AIP Adv. \textbf{5}, 027140 (2015)

\bibitem{spilla_measurement_2014}
S.~Spilla, F.~Hassler, J.~Splettstoesser, New J. Phys. \textbf{16}, 045020
  (2014)

\bibitem{kamp_phase-dependent_2019}
M.~Kamp, B.~Sothmann, Phys. Rev. B \textbf{99}, 045428 (2019)

\bibitem{hajiloo_mesoscopic_2019}
F.~Hajiloo, F.~Hassler, J.~Splettstoesser, Phys. Rev. B \textbf{99}, 235422 (2019)

\bibitem{pershoguba_thermopower_2019}
S.S. Pershoguba, L.I. Glazman, Phys. Rev. B \textbf{99}, 134514 (2019)

\bibitem{guarcello_solitonic_2018}
C.~Guarcello, P.~Solinas, A.~Braggio, F.~Giazotto, Phys. Rev. Appl \textbf{9},
  034014 (2018)

\bibitem{guarcello_phase-coherent_2018}
C.~Guarcello, P.~Solinas, A.~Braggio, F.~Giazotto, Sci. Rep. \textbf{8}, 12287
  (2018)

\bibitem{guarcello_solitonic_2018-1}
C.~Guarcello, P.~Solinas, A.~Braggio, F.~Giazotto, Phys. Rev. B \textbf{98},
  104501 (2018)

\bibitem{bauer_phase-dependent_2019}
A.G. Bauer, B.~Sothmann, Phys. Rev. B \textbf{99}, 214508 (2019)

\bibitem{giazotto_josephson_2012}
F.~Giazotto, M.J. Martínez-Pérez, Nature \textbf{492}, 401 (2012)

\bibitem{martinez-perez_fully_2013}
M.J. Martínez-Pérez, F.~Giazotto, Appl. Phys. Lett. \textbf{102}, 092602
  (2013)

\bibitem{fornieri_nanoscale_2016}
A.~Fornieri, C.~Blanc, R.~Bosisio, S.~D'Ambrosio, F.~Giazotto, Nature Nanotech.
  \textbf{11}, 258 (2016)

\bibitem{fornieri_0-pi_2017}
A.~Fornieri, G.~Timossi, P.~Virtanen, P.~Solinas, F.~Giazotto, Nature Nanotech.
  \textbf{12}, 425 (2017)

\bibitem{giazotto_proposal_2014}
F.~Giazotto, J.W.A. Robinson, J.S. Moodera, F.S. Bergeret, Appl. Phys. Lett.
  \textbf{105}, 062602 (2014)

\bibitem{fornieri_negative_2016}
A.~Fornieri, G.~Timossi, R.~Bosisio, P.~Solinas, F.~Giazotto, Phys. Rev. B
  \textbf{93}, 134508 (2016)

\bibitem{martinez-perez_efficient_2013}
M.J. Martínez-Pérez, F.~Giazotto, Appl. Phys. Lett. \textbf{102}, 182602
  (2013)

\bibitem{giazotto_thermal_2013}
F.~Giazotto, F.S. Bergeret, Appl. Phys. Lett. \textbf{103}, 242602 (2013)

\bibitem{fornieri_normal_2014}
A.~Fornieri, M.J. Martínez-Pérez, F.~Giazotto, Appl. Phys. Lett.
  \textbf{104}, 183108 (2014)

\bibitem{martinez-perez_rectification_2015}
M.J. Martínez-Pérez, A.~Fornieri, F.~Giazotto, Nature Nanotech. \textbf{10},
  303 (2015)

\bibitem{sothmann_high-efficiency_2017}
B.~Sothmann, F.~Giazotto, E.M. Hankiewicz, New J. Phys. \textbf{19}, 023056
  (2017)

\bibitem{guarcello_josephson_2018}
C.~Guarcello, P.~Solinas, A.~Braggio, M.~Di~Ventra, F.~Giazotto, Phys. Rev.
  Appl \textbf{9}, 014021 (2018)

\bibitem{solinas_microwave_2016}
P.~Solinas, R.~Bosisio, F.~Giazotto, Phys. Rev. B \textbf{93}, 224521 (2016)

\bibitem{hofer_autonomous_2016}
P.P. Hofer, M.~Perarnau-Llobet, J.B. Brask, R.~Silva, M.~Huber, N.~Brunner,
  Phys. Rev. B \textbf{94}, 235420 (2016)

\bibitem{vischi_thermodynamic_2019}
F.~Vischi, M.~Carrega, P.~Virtanen, E.~Strambini, A.~Braggio, F.~Giazotto, Sci.
  Rep. \textbf{9}, 3238 (2019)

\bibitem{hwang_nonlinear_2014}
S.Y. Hwang, R.~López, M.~Lee, D.~Sánchez, Phys. Rev. B \textbf{90}, 115301
  (2014)

\bibitem{vannucci_interference-induced_2015}
L.~Vannucci, F.~Ronetti, G.~Dolcetto, M.~Carrega, M.~Sassetti, Phys. Rev. B
  \textbf{92}, 075446 (2015)

\bibitem{ronetti_spin-thermoelectric_2016}
F.~Ronetti, L.~Vannucci, G.~Dolcetto, M.~Carrega, M.~Sassetti, Phys. Rev. B
  \textbf{93}, 165414 (2016)

\bibitem{ronetti_polarized_2017}
F.~Ronetti, M.~Carrega, D.~Ferraro, J.~Rech, T.~Jonckheere, T.~Martin,
  M.~Sassetti, Phys. Rev. B \textbf{95}, 115412 (2017)

\bibitem{roura-bas_enhanced_2018}
P.~Roura-Bas, L.~Arrachea, E.~Fradkin, Phys. Rev. B \textbf{97}, 081104 (2018)

\bibitem{bohling_thermoelectric_2018}
S.~Böhling, G.~Engelhardt, G.~Platero, G.~Schaller, Phys. Rev. B \textbf{98},
  035132 (2018)

\bibitem{roura-bas_helical_2018}
P.~Roura-Bas, L.~Arrachea, E.~Fradkin, Phys. Rev. B \textbf{98}, 195429 (2018)

\bibitem{aharon-steinberg_phenomenological_2019}
A.~Aharon-Steinberg, Y.~Oreg, A.~Stern, Phys. Rev. B \textbf{99}, 041302 (2019)

\bibitem{alicea_new_2012}
J.~Alicea, Rep. Prog. Phys. \textbf{75}, 076501 (2012)

\bibitem{beenakker_search_2013}
C.~Beenakker, Annu. Rev. Cond. Mat. Phys. \textbf{4}, 113 (2013)

\bibitem{aguado_majorana_2017}
R.~Aguado, La Rivista del Nuovo Cimento \textbf{40}, 523 (2017)

\bibitem{fu_superconducting_2008}
L.~Fu, C.L. Kane, Phys. Rev. Lett. \textbf{100}, 096407 (2008)

\bibitem{fu_josephson_2009}
L.~Fu, C.L. Kane, Phys. Rev. B \textbf{79}, 161408 (2009)

\bibitem{rokhinson_fractional_2012}
L.P. Rokhinson, X.~Liu, J.K. Furdyna, Nat. Phys. \textbf{8}, 795 (2012)

\bibitem{wiedenmann_4pi-periodic_2016}
J.~Wiedenmann, E.~Bocquillon, R.S. Deacon, S.~Hartinger, O.~Herrmann, T.M.
  Klapwijk, L.~Maier, C.~Ames, C.~Brüne, C.~Gould et~al., Nat. Commun.
  \textbf{7}, 10303 (2016)

\bibitem{bocquillon_gapless_2017}
E.~Bocquillon, R.S. Deacon, J.~Wiedenmann, P.~Leubner, T.M. Klapwijk,
  C.~Brüne, K.~Ishibashi, H.~Buhmann, L.W. Molenkamp, Nature Nanotech.
  \textbf{12}, 137 (2017)

\bibitem{deacon_josephson_2017}
R.~Deacon, J.~Wiedenmann, E.~Bocquillon, F.~Domínguez, T.~Klapwijk,
  P.~Leubner, C.~Brüne, E.~Hankiewicz, S.~Tarucha, K.~Ishibashi et~al., Phys.
  Rev. X \textbf{7}, 021011 (2017)

\bibitem{li_4pi-periodic_2018}
C.~Li, J.C.d. Boer, B.d. Ronde, S.V. Ramankutty, E.v. Heumen, Y.~Huang, A.d.
  Visser, A.A. Golubov, M.S. Golden, A.~Brinkman, Nat. Mater. p.~1 (2018)

\bibitem{laroche_observation_2019}
D.~Laroche, D.~Bouman, D.J.v. Woerkom, A.~Proutski, C.~Murthy, D.I. Pikulin,
  C.~Nayak, R.J.J.v. Gulik, J.~Nygård, P.~Krogstrup et~al., Nat. Commun.
  \textbf{10}, 245 (2019)

\bibitem{sothmann_fingerprint_2016}
B.~Sothmann, E.M. Hankiewicz, Phys. Rev. B \textbf{94}, 081407(R) (2016)

\bibitem{bours_phase-tunable_2019}
L.~Bours, B.~Sothmann, M.~Carrega, E.~Strambini, A.~Braggio, E.M. Hankiewicz,
  L.W. Molenkamp, F.~Giazotto, Phys. Rev. Appl \textbf{11}, 044073 (2019)

\bibitem{blonder_transition_1982}
G.E. Blonder, M.~Tinkham, T.M. Klapwijk, Phys. Rev. B \textbf{25}, 4515 (1982)

\bibitem{beenakker_josephson_1991}
C.W.J. Beenakker, H.~van Houten, Phys. Rev. Lett. \textbf{66}, 3056 (1991)

\bibitem{beenakker_universal_1991}
C.W.J. Beenakker, Phys. Rev. Lett. \textbf{67}, 3836 (1991)

\bibitem{giazotto_opportunities_2006}
F.~Giazotto, T.T. Heikkilä, A.~Luukanen, A.M. Savin, J.P. Pekola, Rev. Mod.
  Phys. \textbf{78}, 217 (2006)

\bibitem{martinez-perez_quantum_2014}
M.J. Martínez-Pérez, F.~Giazotto, Nat. Commun. \textbf{5}, 3579 (2014)

\bibitem{li_negative_2006}
B.~Li, L.~Wang, G.~Casati, Appl. Phys. Lett. \textbf{88}, 143501 (2006)

\bibitem{paolucci_phase-tunable_2017}
F.~Paolucci, G.~Marchegiani, E.~Strambini, F.~Giazotto, Europhys. Lett.
  \textbf{118}, 68004 (2017)

\bibitem{li_colloquium:_2012}
N.~Li, J.~Ren, L.~Wang, G.~Zhang, P.~Hänggi, B.~Li, Rev. Mod. Phys.
  \textbf{84}, 1045 (2012)

\bibitem{hart_induced_2014}
S.~Hart, H.~Ren, T.~Wagner, P.~Leubner, M.~Mühlbauer, C.~Brüne, H.~Buhmann,
  L.W. Molenkamp, A.~Yacoby, Nat. Phys. \textbf{10}, 638 (2014)

\bibitem{pribiag_edge-mode_2015}
V.S. Pribiag, A.J.A. Beukman, F.~Qu, M.C. Cassidy, C.~Charpentier,
  W.~Wegscheider, L.P. Kouwenhoven, Nature Nanotech. \textbf{10}, 593 (2015)

\bibitem{amet_supercurrent_2016}
F.~Amet, C.T. Ke, I.V. Borzenets, J.~Wang, K.~Watanabe, T.~Taniguchi, R.S.
  Deacon, M.~Yamamoto, Y.~Bomze, S.~Tarucha et~al., Science \textbf{352}, 966
  (2016)

\bibitem{giazotto_coherent_2013}
F.~Giazotto, M.J. Martínez-Pérez, P.~Solinas, Phys. Rev. B \textbf{88},
  094506 (2013)

\bibitem{timossi_phase-tunable_2018}
G.F. Timossi, A.~Fornieri, F.~Paolucci, C.~Puglia, F.~Giazotto, Nano Lett.
  \textbf{18}, 1764 (2018)

\bibitem{paolucci_phase-tunable_2018}
F.~Paolucci, G.~Marchegiani, E.~Strambini, F.~Giazotto, Phys. Rev. Appl
  \textbf{10}, 024003 (2018)

\bibitem{lee_inducing_2017}
G.H. Lee, K.F. Huang, D.K. Efetov, D.S. Wei, S.~Hart, T.~Taniguchi,
  K.~Watanabe, A.~Yacoby, P.~Kim, Nat. Phys. \textbf{advance online
  publication} (2017)

\bibitem{guiducci_toward_2019}
S.~Guiducci, M.~Carrega, G.~Biasiol, L.~Sorba, F.~Beltram, S.~Heun, physica
  status solidi (RRL) – Rapid Research Letters \textbf{13}, 1800222 (2019)

\bibitem{sochnikov_nonsinusoidal_2015}
I.~Sochnikov, L.~Maier, C.A. Watson, J.R. Kirtley, C.~Gould, G.~Tkachov, E.M.
  Hankiewicz, C.~Brüne, H.~Buhmann, L.W. Molenkamp et~al., Phys. Rev. Lett.
  \textbf{114}, 066801 (2015)

\bibitem{fornieri_evidence_2019}
A.~Fornieri, A.M. Whiticar, F.~Setiawan, E.~Portolés, A.C.C. Drachmann,
  A.~Keselman, S.~Gronin, C.~Thomas, T.~Wang, R.~Kallaher et~al., Nature
  \textbf{569}, 89 (2019)

\bibitem{ren_topological_2019}
H.~Ren, F.~Pientka, S.~Hart, A.T. Pierce, M.~Kosowsky, L.~Lunczer,
  R.~Schlereth, B.~Scharf, E.M. Hankiewicz, L.W. Molenkamp et~al., Nature
  \textbf{569}, 93 (2019)

\bibitem{tkachov_helical_2013}
G.~Tkachov, E.M. Hankiewicz, Phys. Rev. B \textbf{88}, 075401 (2013)

\bibitem{tkachov_magnetic_2004}
G.~Tkachov, V.I. Fal’ko, Phys. Rev. B \textbf{69}, 092503 (2004)

\bibitem{tkachov_quantum_2015}
G.~Tkachov, P.~Burset, B.~Trauzettel, E.M. Hankiewicz, Phys. Rev. B
  \textbf{92}, 045408 (2015)

\bibitem{pikulin_disorder_2014}
D.I. Pikulin, T.~Hyart, S.~Mi, J.~Tworzydło, M.~Wimmer, C.W.J. Beenakker,
  Phys. Rev. B \textbf{89}, 161403 (2014)

\bibitem{tkachov_ballistic_2010}
G.~Tkachov, E.M. Hankiewicz, Phys. Rev. Lett. \textbf{104}, 166803 (2010)

\bibitem{bours_topological_2018}
L.~Bours, B.~Sothmann, M.~Carrega, E.~Strambini, E.M. Hankiewicz, L.W.
  Molenkamp, F.~Giazotto, Phys. Rev. Appl \textbf{10}, 014027 (2018)

\bibitem{giazotto_superconducting_2010}
F.~Giazotto, J.T. Peltonen, M.~Meschke, J.P. Pekola, Nat. Phys. \textbf{6}, 254
  (2010)

\bibitem{strambini_omega-SQUIPT_2016}
E.~Strambini, S.~D'Ambrosio, F.~Vischi, F.S. Bergeret, Y.V. Nazarov,
  F.~Giazotto, Nature Nanotech. \textbf{11}, 1055 (2016)

\bibitem{vischi_coherent_2017}
F.~Vischi, M.~Carrega, E.~Strambini, S.~D'Ambrosio, F.S. Bergeret, Y.V.
  Nazarov, F.~Giazotto, Phys. Rev. B \textbf{95}, 054504 (2017)

\bibitem{hwang_phase-coherent_2018}
S.Y. Hwang, F.~Giazotto, B.~Sothmann, Phys. Rev. Appl \textbf{10}, 044062
  (2018)

\bibitem{riwar_multi-terminal_2016}
R.P. Riwar, M.~Houzet, J.S. Meyer, Y.V. Nazarov, Nat. Commun. \textbf{7}, 11167
  (2016)

\bibitem{eriksson_topological_2017}
E.~Eriksson, R.P. Riwar, M.~Houzet, J.S. Meyer, Y.V. Nazarov, Phys. Rev. B
  \textbf{95}, 075417 (2017)

\bibitem{meyer_nontrivial_2017}
J.S. Meyer, M.~Houzet, Phys. Rev. Lett. \textbf{119}, 136807 (2017)

\bibitem{xie_topological_2017}
H.Y. Xie, M.G. Vavilov, A.~Levchenko, Phys. Rev. B \textbf{96}, 161406 (2017)

\bibitem{xie_weyl_2018}
H.Y. Xie, M.G. Vavilov, A.~Levchenko, Phys. Rev. B \textbf{97}, 035443 (2018)

\bibitem{hogan_ferromagnetic_1953}
C.L. Hogan, Rev. Mod. Phys. \textbf{25}, 253 (1953)

\bibitem{viola_hall_2014}
G.~Viola, D.P. DiVincenzo, Phys. Rev. X \textbf{4}, 021019 (2014)

\bibitem{mahoney_-chip_2017}
A.~Mahoney, J.~Colless, S.~Pauka, J.~Hornibrook, J.~Watson, G.~Gardner,
  M.~Manfra, A.~Doherty, D.~Reilly, Phys. Rev. X \textbf{7}, 011007 (2017)

\end{thebibliography}

\end{document}